**C<>DE** UNIVERSITY OF APPLIED SCIENCES

Faculty of Product Management

## Bachelor's Thesis

# The impact of no-code on digital product development.

Simon Heuschkel

Date of submission: 05.12.2022

Semester: Fall 2022

First Assessor: Florian Grote

Second Assessor: Swantje Quoos

# Declaration

I hereby declare that the work titled "**The impact of no-code on digital product development**" was written independently, as I am the sole author. I did not use any other assistance or sources other than those mentioned. All references taken literally or correspondingly from published and unpublished writings are marked as such. I am aware that any content used from the internet must have been acknowledged and added as an electronically stored resource.

This work was prepared to be presented and evaluated as a Bachelor's Thesis. It was not submitted either in its entirety or parts thereof to any other examination authority.

Date:                              Signature

# Abstract

Low- and no-code platforms (LCNC) have become more popular than ever (Kulkarni, 2021), with low-code broadly adopted to optimise internal business processes. Increasingly, startups build their primary software product using no-code platforms (Palios, 2022). This paper explores why entrepreneurs choose no-code platforms to build, launch and scale a software product, what benefits and limitations no-code has, and why they might transition to custom-developed solutions later. Ten semi-structured interviews with successful projects and no-code startup founders were conducted. The results show that speed, cost savings and the lack of coding knowledge are the primary reasons entrepreneurs choose no-code initially. Challenges are diverse and depend on the no-code platform, the maker's skill and the product. The impact of no-code on established product development frameworks and the maker's role are discussed.

# Table of Content

# 1 List of Abbreviations and Definitions

- **AI:** Artificial intelligence.
- **Citizen Developer (CD):** A CD describes the person using LCNC platforms, usually having no or little formal programming experience. This person can be an employee using LCNC to improve business processes (outside of the IT department) but can also build software for other (personal) use cases.
- **IoT:** Internet of things.
- **LCNC:** Low-code, no-code.
- **Low-code:** Platforms that heavily reduce development time and cost by reducing the amount of "coding" to a minimum. Citizen developers usually use them to optimise large enterprise processes.
- **Maker(s):** The individuals that were interviewed for this study - the interviewees. A "creator" or someone who "turns ideas into reality" (Nira Team, n.d., p. 3)**.**
- **No-code (platforms):** Definition used in this research: "Platforms that allow non-technical people to build complex software products, such as custom mobile apps, web apps, or marketplaces, without writing code."
- **No-code startup:** This paper defines it as: "Projects or startups that built a successful, custom software product using no-code solutions". It includes all of the projects and startups interviewed in this paper and does not mean the no-code platform providers.
- **PM:** Product Management.
- **Rapid application development (RAD)**: RAD is a framework developed by James Martin and is also a general term used to describe adaptive software development practices. Compared to Agile, RAD shares similar values like flexibility, shorter delivery times and high customer interaction but is focused more on complete prototypes rather than breaking down features to be delivered in a sprint.

# 2 List of Figures

# 3 Introduction

Since the introduction of the personal computer, using and creating software has become increasingly accessible.

From the introduction of graphical user interfaces (Isaacson, 2012, p. 124), the first no-code website builders like WordPress (Palios, 2022), to the increasing use of low and no-code (LCNC) platforms in large enterprises. With more applications needed than ever and a shortage of developers (Lebens et al., 2021), LCNC platforms see a steady increase in adoption.

Since the 2000s, non-technical people outside the IT department have used low-code platforms to automate business processes by building internal business applications (Citizen Development) (Luo et al., 2021).

In recent years, no-code platforms enabled non-technical individuals to develop custom software without writing code (Palios, 2022).

This paper explores the impact of no-code on digital product development. Using ten semi-structured interviews, it analyses successful projects and startups that used no-code platforms to build their custom software product without writing code. These 'no-code startups" were selected using purposive sampling based on their primary no-code tech stack. Three no-code platforms, Bubble.io, Softr.io and Sharetribe.com, were selected for this research.

The aim is to answer why no-code startups build, launch and scale software products using no-code, the benefits and limitations of doing so, and why (if at all) they might transition to custom-code later on.

## 3.1 Research Question and Hypothesis

Two research questions and two underlying hypotheses were chosen to explore the impact of no-code on digital product development.

### 3.1.1 Research Question

- Why do software startups build, launch and scale their product using no-code solutions? Including the decision-making process, requirements and long-term product strategy.

- What benefits and limitations do no-code platforms currently have, and why (if at all) do software startups choose to transition to a custom solution?

## 3.1.2 Hypothesis

- (1) Software-only startups using no-code for their initial product (minimal viable product (MVP)) have significant advantages such as faster time to market, iterations and feedback cycles, earlier user growth and saved cost.
- (2) As the startup and its product(s) grow, no-code platforms become more limiting and potentially need to be replaced with a custom solution.

# 4 Literature Review and Context

## 4.1 History of No-code

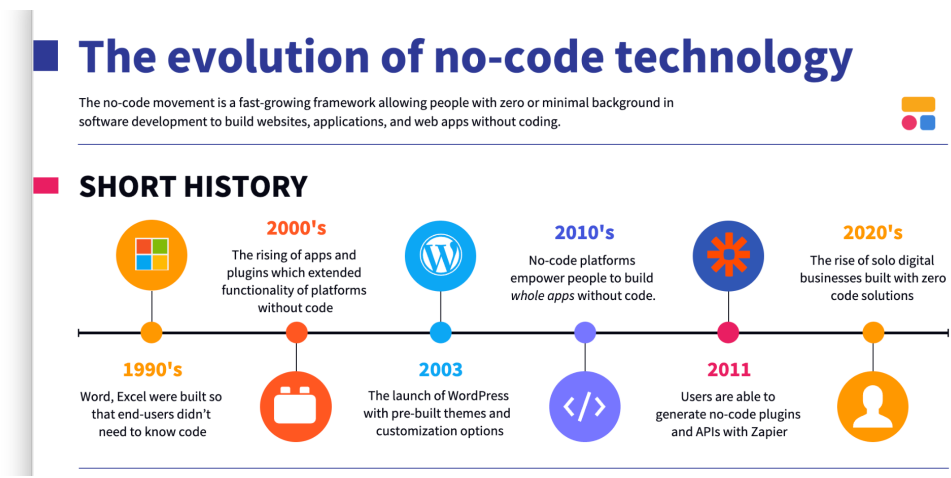

Figure 1:

*The Evolution of No-code Technology (Palios, 2022)*

The following section summarises the work of (Jonathan, n.d.), (Palios, 2022) and (History of the No-Code Movement, n.d.). For a detailed history timeline, refer to Appendix 11.2.

Looking at the broad history of technology, we find multiple examples of democratisation. From the printing press in 1440 to typewriters and digital technologies that allow anyone to publish anything today. The creation of software is no different.

The first computers in the 1970s needed technical knowledge and used a command-line interface. In 1975 researchers from Xerox PARC, among them, computer pioneer Alan Key demoed the first graphical user interface (GUI). It laid the foundation of modern software and inspired Steve Jobs and Bill Gates to equip personal computers with a GUI. For the first time, non-technical users could use a computer and interact with software.

**All-in-one platforms and single-function software (the 1990s):** By 1990, innovators like Microsoft and Adobe had built all-in-one programs like Word, Excel and Adobe Photoshop that massively democratised how ordinary people could use software.

**Extended functionality, plug-ins, and app ecosystems (the 2000s):** With the new millennial starting, software as a service (SaaS), Email automation, low-code and basic website builders inaugurated. Suddenly, people could build and publish websites and send stylised rich text emails. Mid-century, the idea of plug-ins and app stores to easily extend the functionality of a website or SaaS without code was born. Developing apps or plug-ins became lucrative for developers.

With low-code, citizen developers could build internal enterprise tools to optimise business processes. It still required the end user to use some coding to build an application. Forrester (Richardson & R. Rymer, 2014) first defined low-code as: "Platforms that enable rapid application delivery with a minimum of hand-coding, and quick setup and deployment, for systems of engagement" and added attributes such as "visual development, automatic configuration and deployment, or user interface transcoding" with the main benefits being "cost" and "speed of development."

**Build custom apps without code (the 2010s - present):** Until now, building custom software still required some software engineering knowledge. No-code platforms launched and enabled anyone to build. "No-code development is the term given to a variety of tools that help people build software without code" (Palios, 2022). End users do not require any coding knowledge to build something totally custom. Technically, this includes tools that previously required software knowledge, such as email marketing tools and website builders.

**Future of no-code:** No-code has continuously democratised technology. There are three likely frontiers for no-code: Artificial intelligence (AI), Web3 and blockchain technology, and the Internet of Things (IoT). These areas are cutting-edge today and require software engineering knowledge to train and use AI models or build IoT infrastructure. It will only be a matter of time until even these technologies can be used and built by anyone and new frontiers for software developers arise.

## 4.2 LCNC Market Overview

The global low-code platform market revenue was almost 13 billion U.S. dollars in 2020 and is expected to reach approximately 65 billion U.S. dollars in 2027, with a CAGR of 26.1% over this period (Vailshery, 2022). Leading low-code platform providers raised an estimated $807 million and are valued at over $361 billion (Kulkarni, 2021). 70% of new applications developed by organisations will use LCNC technologies by 2025, up from less than 25% in 2020 (Govekar, 2021). The no-code adoption grew during the corona pandemic, and most users plan to use it for personal projects (Zapier Editorial Team, 2022).

Two trends can explain the growth: The number of digital applications and services being built is exploding. Between 2018 and 2023, more than 500 million apps will be created, more than the previous 40 years combined (Gens, 2019).

At the same time, there is a need for more skilled software engineers. According to McKinsey, IT is the second-largest area for businesses to address potential skill gaps after data science (Agrawal et al., 2020)

## 4.3 Relevant No-code Platforms

The selected platforms are used by makers in this paper and are market leaders in their respective niches. (Bubble Group, Inc., 2020; Nira Team, n.d.; Zapier Editorial Team, 2022)

Funding data was obtained from Crunchbase (crunchbase.com) on Nov. 20th. 2022.

A detailed list of leading platforms is available in Appendix 11.3.

### 4.3.1 Bubble (Bubble.io, n.d.)

Founded in 2012 | Funding: $106 million | 2.120.512 users

# The best way to build
## SaaS apps
# without code

Building tech is slow and expensive. Bubble is the most powerful no-code platform for creating digital products. Build better and faster.

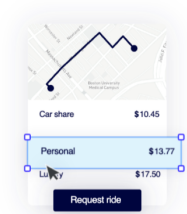

**Get started for free**    Join **2,120,512** Bubblers today.

Figure 2: *Bubble Website Screenshot (Bubble.io, n.d.)*

"Bubble lets you create interactive, multi-user apps for desktop and mobile web browsers, including all the features you need to build a site like Facebook or Airbnb. Build logic and manage a database with our intuitive, fully customisable platform" (Bubble.io, n.d.). Bubble has a visual interface separated into multiple parts bringing together the final product: design, workflows, data, styles, plug-ins, version control and settings. It built up a thriving community over the past years. They offer a template marketplace (900+) that allows anyone to start with the majority of work already done, a forum to ask questions, 60+ official plug-ins like Stipe (stripe.com), several hundred community plug-ins, and a network of Bubble experts and agencies. Most agree that Bubble is best suited to build custom, scalable solutions.

Pricing: They offer a Bubble-branded free plan without a custom domain and paid plans between $29-$529 / month.

### 4.3.2 Sharetribe (Sharetribe, n.d.)

Founded in 2011 | Funding: $2,4 million | 1k+ marketplaces created.

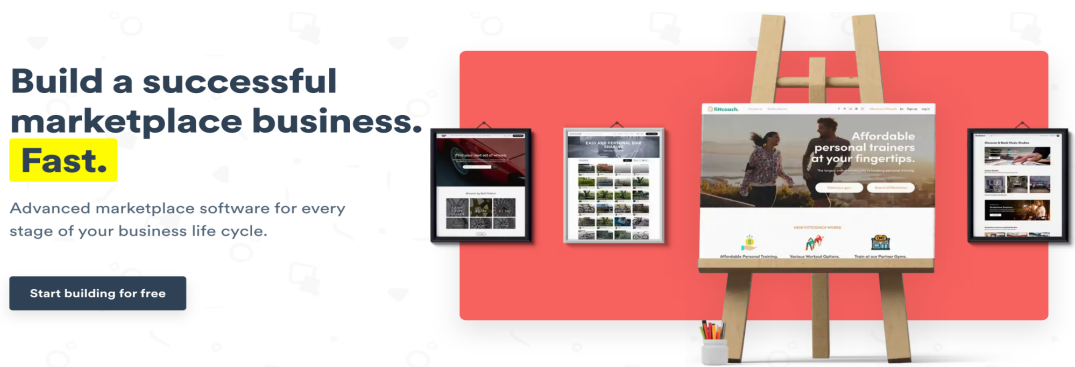

Figure 3: *Sharetribe website screenshot (Sharetribe, n.d.)*

"Advanced marketplace software for every stage of your business life cycle" (Sharetribe, n.d.). Sharetribe offers two plans: Go and Flex. Go is the plug-and-play solution that allows users to "build a fully functional marketplace [like Airbnb or Fiverr] in one day". Out of the box, it offers users to "create profiles, list their offering, interact with each other, and make online payments" and offers builders tracking and analytic features in a dashboard, user access control and customised branding (Sharetribe, n.d.).

Sharetribe Flex offers unique low-code features. "Flex lets you custom-develop your marketplace freely – at a fraction of the cost of building and hosting your platform" (Sharetribe, n.d.). It offers everything a basic marketplace needs and allows to change or add custom-coded features. Builders can use well-documented APIs, export the entire code and optionally host it on their server. Sharetribe allows makers to quickly launch advanced marketplace (especially if they do not know Bubble).

They offer a free 30-day trial which is enough to launch a marketplace.

Pricing: Prices range between $99-$299/ month on the Go plan for $100-$100,000 users and $299 / month plus dynamic pricing based on transaction volume for Flex. In both plans, the Stripe gateway fees are charged on top.

### 4.3.3 Softr (Softr.Io, n.d.)

Founded in 2019 | Funding: $15 million | 80.000 users

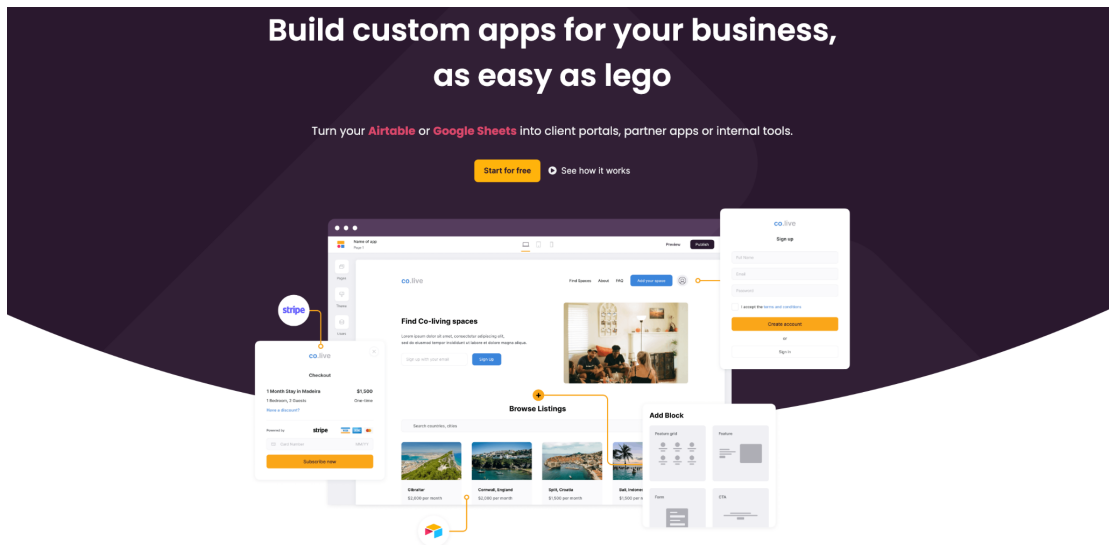

Figure 4: *Softr Website Screenshot (Softr.io, n.d.)*

"Softr is the easiest, fastest way to build a professional web app on Airtable [and Google Sheets]. Zero learning curve" (Softr.Io, n.d.). With Softr, users can build client portals, internal tools, marketplaces, online communities, resource directories and websites. It acts as a front-end to existing Airtable or Google Sheet data. Softr is unique due to its 100+ pre-built blocks and native integration like Mailchimp (mailchimp.com), Stripe (stripe.com) and analytics tools. Additionally, it provides user management, mobile progressive web app functionality and free custom domains.

If the data already exists, users can launch within hours, and with their feature-rich free plan, builders can launch complete products.

Pricing: Prices range from $29-$199 based on the platform's number of members and rows in the database.

### 4.3.4 Airtable (Airtable.com, n.d.)

Founded in 2013 | Funding: $1.4 billion | 300.000 customers

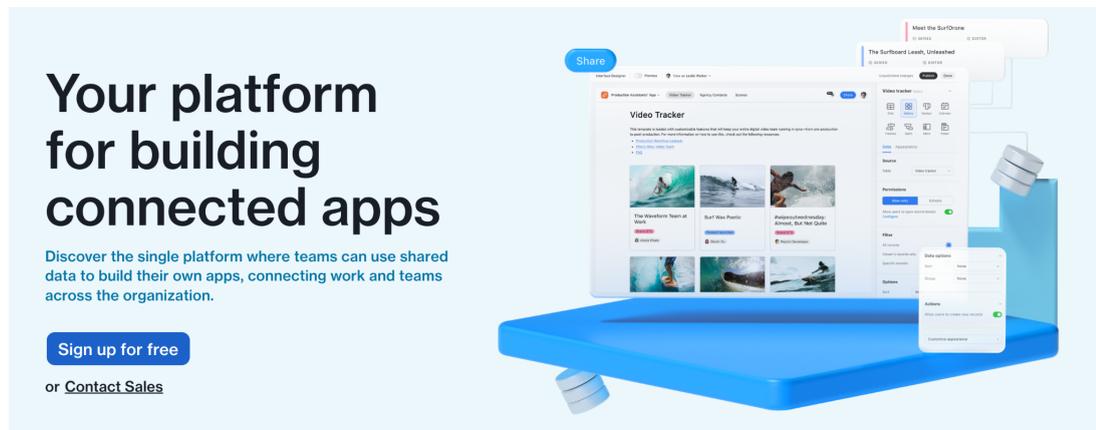

Figure 5: *Airtable website screenshot (Airtable.com, n.d.)*

"Your platform for building connected apps. Discover the single platform where teams can use shared data to build their own apps, connecting work and teams across the organisation" (Airtable.Com, n.d.).

Airtable started as a visual database editor. The fields in a table are similar to cells in a spreadsheet but have types such as 'checkbox', 'phone number' or 'drop-down list, and can reference file attachments like images. Users can create a database, set up column types, add records (rows), link tables to one another, collaborate, sort records and publish views to external websites. Information is organised in a base which "is a collection of related tables, often reflecting a single project, process, or workflow that you collaborate on with your team" (Airtable.Com, n.d.).

It has become the database tool of choice for most no-code tools and makers, with its own marketplace of integrations and plug-ins. Many businesses use it to unify data in one central place, collect information with built-in forms, and display data visually using the "interface designer". Airtable is ideal for data-heavy projects.

Pricing: Their free plan offers 1,200 records per base, a powerful free plan. Their non-enterprise paid plans are $10 for 5,000 records and $20 for 50,000 records.

### 4.3.5 Zapier (Zapier.com, n.d.)

Founded in 2011 | Funding: $1.4 million | 2.2 million business customers

Figure 6: *Zapier Website Screenshot* (*Zapier.Com*, 2022)

"From side hustlers to enterprise leaders, Zapier connects your work apps, so you get more focus and less frustration" (Zapier.com, n.d.). Zapier is a visual API connector tool to connect and integrate 5,000+ apps. Users can create multi-step workflows, add conditional logic and format data between apps. An example can look like this: If a new deal is created in Hubspot, inform the team on Slack and create a new folder in Google Drive. More and more tools try to bring information from different apps together, and Zapier is the "backend" for this. Zapier is the most popular tool to connect apps and automate workflows.

Pricing: Zapier offers a free plan that includes five Zaps (automation workflows) and 100 tasks/month. Paid plans start from $19.99 (20 Zaps, 750 tasks/month) to $799 (unlimited Zaps, 100,000 tasks/month).

## 4.4 Review of No-code Literature

Most available literature revolves around low-code use cases and their characteristics. In contrast, available no-code literature is rare and primarily initiated by the platform providers. **Characteristics and Challenges of low-code:** Two studies analysed LCNC discussions using online Q&A platforms like Stack Overflow (SO) and Reddit.

Luo et al. (2021) analysed 73 SO and 228 Reddit posts to uncover the characteristics and challenges of low-code development platforms from the practitioner's perspective. They found that low-code platforms are especially needed to automate processes and workflows and may provide a graphical user interface with drag and drop, out-of-the-box components (e.g., APIs connections), which makes them easy to learn and use and speeds up the development. According to the research, "practitioners have conflicting views on the advantages and disadvantages" (Luo et al., 2021, p. 1).

Similar research using only SO analysed around 5,000 posts to discover developers' challenges during their adoption of low-code platforms (Alamin et al., 2021) found that 40% of the questions were about customisation, with dynamic event handling being the most popular and most challenging topic. 75% of the questions remain unanswered and are related to the project's development phase. They conclude that more documentation must be required to adopt low-code platforms smoothly.

**Reasons for and against low-code:** Alsaadi et al. (2021) surveyed 49 developers to find out what factors attract or prevent developers from using low-code solutions. They found the most significant advantage of low-code was reduced development time (95% of the sample), and the minimum coding required made development more straightforward. Disadvantages mentioned by developers who used low-code were customisation of built-in components and lack of knowledge about these platforms. Developers who did not yet use low-code platforms mentioned scalability as their main reason not to use it, which aligns with (Alamin et al., 2021).

**Leading low-code platforms:** Kulkarni (2021) analysed various low-code market reports. He found that low-code in the business context is here to stay and integrates well into existing development frameworks like rapid application development (RAD) and Agile. Leading low-code providers are identified based on industry reports like Forrester and Gartner. Namely Appian, Mendix, Microsoft Power Apps, Outsystems, Pegasystems, Salesforce and ServiceNow. Together they received funding of $807 million with a market cap of $361 billion.

**Low-code features:** Sahay et al. (2020) analysed eight low-code platforms based on a predefined set of features and built a sample application to uncover limitations. They found a set of base features: a visual editor with pre-built components, interoperability support (plug-ins and APIs to other platforms), security controls, logic builder, dynamic scalability, and deployment support. Challenges they experienced included: The lack of standards (interoperability) between platforms, extensibility outside the core features requiring extensive coding, a high learning curve where software development knowledge is helpful, and the lack of information on how scalability is handled within these platforms.

**Entrepreneurship:** Alamin et al. (2021) analysed 400 VC-backed e-commerce startups and compared the growth trajectory between Shopify-based (no-code) and custom-developed startups during the 2010s. They found that Shopify-based startups start with fewer resources and achieve similar successful exits. "The value created per employee, and cash-on-cash return for investors, place Shopify-based startups on par with their peers" (Dushnitsky & Stroube, 2021).

**Security:** Oltrogge et al. (2018) analysed over two million free Android apps to assess the security impact of online app generators (OAG - LCNC app builders). 11.1% of these apps were created using OAGs and found several well-known security issues (e.g. code injection) in some of the boilerplate code. Through the "black box development approach, citizen developers are unaware of these hidden problems that ultimately put the end-user's sensitive data and privacy at risk and violate the user's trust assumption" (Oltrogge et al., 2018, p. 1).

**Low-code testing:** Khorram et al. (2020) analysed five leading low-code platforms regarding their testing abilities. They found that there are no general low-code testing frameworks in place. Additionally, citizen developers are still highly dependent on third-party testing tools they cannot use.

**Supply chain management:** Bhattacharyya and Kumar (2021) explored the application of low-code platforms to digitalise supply chains. They found that low-code can be used in the supply chain and could reduce the dependency on software engineers. Especially for SMEs, the cost savings and ease of use could "level the playing field against large organisations" (Bhattacharyya & Kumar, 2021).

**Manufacturing:** Sanchis et al. (2020) did a context analysis of low-code platforms focusing on the manufacturing industry. They recognise the lack of research in this niche but highlight that LCNC platforms for virtual factory open operating systems (vf-OS) are emerging.

No-code-specific research in the context of entrepreneurship and startups is rare. However, three sources could be identified, and it is assumed that their target audience overlaps with this study's participants.

**No-code consensus:** A Bubble Group, Inc. (2020) survey with a total of 741 respondents (from 1414 people reached) displayed to Bubble users, and a partner network found:

- Most no-coders work in startups or are self-employed, mostly in bootstrapped companies.

- Respondents said no-code is faster, more affordable, and easier to use than traditional code.
- The main reasons for no-code are a faster build, building without engineers, easier to use, control over the product, and affordability.
- The best products are Bubble, Airtable, Zapier, WordPress, and Shopify.
- The average no-coder is in their mid-thirties and works in an average median company size of three employees/colleagues in a 2-3-year-old business.

**The state of makers:** An access control startup surveyed 260 makers (defined as "creator" or someone who "turns ideas into reality) and found that only 19% of makers used no-code (Nira Team, n.d., p. 3). People who used no-code used the following tools: Zapier, GSuite, Airtable, Shopify, Webflow, Coda, Bubble, and Carrd. They found that challenges for makers are keeping up productivity, marketing and sales, "making something useful", profitability, shipping fast and limited resources. They predict a "flood of new makers" that will have more time to focus on marketing and sales as product development will become more accessible and faster.

      **The rise of no-code:** Zapier (2022) researched the use of no-code and found that: Most 'no-coders' started using no-code during the corona pandemic, think their company could grow faster, use no-code for their personal projects, and plan to use it more in the future.

The literature shows that enterprises have used low-code for many years to help optimise various business processes.

No-code is emerging, carving out its use cases focusing on startups and makers.

# 5 Methodology and Study Design

## 5.1 Selection of Methods

The study uses a mixed method approach of qualitative semi-structured interviews combined with a quantitative rating scale survey.

The goal of the semi-structured interviews was to gather in-depth contextual information about the no-code startups and better understand their decisions regarding no-code. Additionally, available LCNC research primarily uses quantitative methods.

The rating scale survey gathers quantitative insights to weigh interview answers and backcheck established no-code views.

## 5.2 Selection of No-code Platforms and Samples

For both the no-code platform and the sample, purposive sampling was used.

**Selection of no-code platforms:** The goal was to find three no-code platforms (as defined for this research) with successful no-code startups using these platforms. Significant traction was used to define success and included page views, users, funding amount or revenue, depending on the project.

Based on this definition, platforms like Bubble, Sharetribe, Airtable, Glide (glideapps.com), Bravo Studio (bravostudio.app), or Adalo (adalo.com) are included, and email automation tools, website/eCommerce builders like Shopify (shopify.com), or Wix (wix.com) and enterprise-grade low-code platforms like Outsystems (outsystems.com) or Mendix (mendix.com) are excluded, as they are primarily used to build internal applications for existing businesses.

Over 40 resources were analysed, such as the "showcase" sections of no-code platforms, blog articles and no-code project directories.

Bubble.io, Softr, and Sharetribe best fit the criteria and were chosen.

**Selection of samples:** In total, 112 no-code startups were analysed, of which 83 used the selected no-code platforms. Each no-code startup was rated (1-10) based on their traction and received a note with essential details. Information was obtained from their website, showcase articles on the no-code platform or other information sources like blog posts or Crunchbase.

Only no-code startups with the two highest ratings (53 items) were contacted and interviewed on a first-come-first-serve basis. The final sample consisted of three Bubble, three Sharetribe and four Softr startups. A one-page summary of each interview can be found in Appendix 11.1.

For the purpose of this publication, some startup and maker names were anonymised and replaced with pseudo-names. They are: Interiorly (by Shrikar), Bike2loop (by Mark), ClimateCollabEU (by Tom), Talent Tap (by Fernando), Connectolingo (by Benjamin), Locumify (by Pranav)

## 5.3 Qualitative Semi-Structured Interviews

### 5.3.1 Question Design

The interview questions are based on the two research questions (RQ) and contribute to either of the RQs, as shown in Figure 7.

| Qualitative Interview Questions aligned to RQs | |
|---|---|
| **Research Question 1 (RQ1)** | **Research Question 2 (RQ2)** |
| 1.) Why do software startups build, launch and scale their product on top of a no-code solution? 2.) This includes the decision-making process, requirements and long-term product strategy. | 1.) What benefits and limitations do no-code platforms currently have? 2.) What (if at all) do software startups choose to transition to a custom solution? |
| **Background & Context** | **Early days: Zero to one.** |
| Could you share a bit about you? Whats your background, age? | What challenges did you encounter while building your initial product with no-code? |
| How did you get into no code? | What benefits did you encounter while building your initial product with no-code? |
| What no code tools do you know? | |
| Describe your startup in a few sentences. | |
| What type of project is that ? Side Hustle/Startup | **Later days: One to infinity (Growth).** |
| What's your team size? | What challenges did or do you currently have during growth in regards to the chosen no code platform? |
| Whats your Revenue / traction | What benefits did/did you encounter during startup growth in regards to the chosen no code platform? |
| | What changes did you perform or features did you add after your initial product launch? What role did no code play? |
| **Early days: Zero to one (Launch)** | Have you thought about re-building your product with a custom code solution? If yes, why. If not, why? |
| What were your initial product requirements? | Did you switch or modify your tech stack after launch? If yes, why? |
| What where your initial requirements for a no code platform? | Looking back what would you change? |
| Which no code stack / platform did you use? | Looking back what would you keep? |
| Why did you choose to build and launch your product using the above no code platform(s) ? | |
| Did you try other solutions before? Why did you decide against them? | |
| Why did you decide against building a custom coded solution? | |
| What was/is your long-term product strategy? | |
| What was your time to market? How was that influenced by using no code? | |

Figure 7: *Qualitative Interview Questions Aligned to RQs*

An interview question blueprint was prepared and organised to chronologically cover the project journey and retrieve information about the makers' experience (See complete interview questions blueprint in Appendix 11.4).

The interview questions are designed to capture the thinking process for every step of the journey. Some questions might seem repetitive. However, they cover different points in time and may provide unique insights.

The deep dive into what they changed after launch aims to find specific examples of benefits and challenges they might miss in the dedicated question. The same is true for the reflection question (keep or change).

## 5.3.2 Execution

All interviews were conducted and recorded via the video conferencing platform Zoom (zoom.us) and had a time limit of 30-60 minutes. At the end of each interview, the interviewee completed the rating scale survey. Each recording was uploaded to the transcription service Otter (otter.ai) and manually edited afterwards. The transcribed interviews are the basis for the analysis.

## 5.3.3 Analysis

Excel was chosen for the analysis to avoid platform lock-in. Every insight can be traced back to the origin of the transcript. The raw data and analysis are available in Appendix 9.

**Independent interview analysis:** Each interview was first analysed independently. Each interview question was organised along the x-axis, and each no-code startup, the learnings and the corresponding source along the y-axis (Fig. 8). In the case where the interviewee answered multiple questions at once, answers were collected from the transcript without directly stating the question.

Figure 8: *Independent interview analysis*

**Meta-analysis:** For the meta-analysis, the interviews were regrouped to show learnings by interview, no-code platform, and allow for summaries with different granularity. Along the x-axis, each no-code startup was grouped based on the platform used. The source row was hidden. On the y-axis, each interview question was listed along with the learnings (Fig. 9).

| | Interiorly | 100 Days of no code | AIDE Hulp | ClimateCollabEU | Connectolingo | Locumify |
|---|---|---|---|---|---|---|
| Describe your startup in a few sentences. | A collection of semi-customized Interior design marketing content. Pictures, tags, captions, and canva files that saves time when planning content. For english speaking audience, and design businesses of not more than 3 million revenue a year | - It's an online community, empowering creators and entrepreneurs to leverage NoCo tools so that they can bring their ideas to life. - 100 days of no code (30 min email course) - Community (slack) - Bootcamp - In person, online live bootcamp for accelerated learning. | - A marketplace to ask for and receive help from people. - Crisis response page for floods that happened in Belgium | - platform for anyone who wants to do anything, in Europe related to climate change, or climate tech. - Job platform - Investment insights platform - Resources - Online community | -Interactie Language learning app - Many different activities and bite-sized lessons - Focused on Hebrew and Arabic. - "Hebrew for French speakers, Hebrew for English, Hebrew for Arabic speakers. And you can learn Arabic for English speakers" | -Job matching platform for doctors - They bring together all the different health care providers, recruitment agencies and doctors who are looking for locum jobs or short term work. - Freelance doctors (locum work) has many benefits such as higher pay. - Because there is much more demand for doctors doctors can choose where they want to work, which is why their platform works. |
| Summary | - 6x: Marketplace (Bubble, Softr and Sharetribe) - 2x: Community (with Directory based products) (Softr) - 1x: Course based Web App (Bubble + Articulate 360) - 1x: Info/directory based product (Softr) | | | | | |
| Main or Side Hustle ? | Side hustle | Main, full time work. Also Runs: No code for kids - education in UK schools | - Side hustle, volunteered during flood crisis | - Full time job, started making money past week. | Full time, got funded by bigest news media group in Isreal. | - Main job, full time for the past 4 years. |
| Summary | - Side Hustle 3x - Main / Full time Job 7x | | | | | |
| Team Size | 1 + freelancers for some jobs. | in a team of three. | - Two Freelance friends | - One person - Onboarded head of product and communications this week. | 9 People full time | - Team of 9 (from website) |

Figure 9: *Meta-Analysis*

**Summary:** The summary links the individual interview question insights to each RQs to find answers independently from a logical project timeline (Fig. 10).

| Impact of no-code on digital product development | | | |
|---|---|---|---|
| **Research Question 1** 1.) Why do software startups build, launch and scale their product on top of a no-code solution? 2.) This includes the decision-making process, requirements and long-term product strategy. | | **Research Question 2** 1.) What benefits and limitations do no-code platforms currently have? 2.) Why (if at all) do software startups choose to transition to a custom solution?" | |
| **Background and context** | | | |
| Tell us a bit about you? Whats your background, | - Most are in their thirties... - All of them have a marketing or business background, one is a Doctor. | | |
| How did you get into no code? | Three ways to get into no code: - Make process more efficient in existing company. - Accidentally found it online in a YT comment or Article - Actively researched cheap and quick alternatives to code. | | |
| What no code tools do you know? | Everyone know Bubble.io & Zapier Many know Airtable and use Stripe. | | |
| Main or Side Hustle ? | - Side Hustle 3x - Main / Full time Job 7x | | |
| Team Size? | Either small teams of 1-4 people or established teams of 9 or 50 full time people. | | |
| Traction | All of the project have significant traction in terms of users, engagement, funding, and or revenue. Especially Bubble and Sharetribe Apps. Softr apps reserved allot of engagement, but monetisation was less developed. Probably due to the young | | |
| **Product and no-code requirements** | | **Benefits** | |
| Initial product requirements? | As most of the product interviewed are marketplaces or directory based it is no wonder, that creating listings, accounts, and payments and browsing, filtering listings are the top requirements. | Benefits for MVP | - Speed: 6x - Bubble: Predefined, UX, blocks and templates (softr) 2x - Bubble: Visual interface, ""no translation"" of requirements, no bugs, front/back end connection removed common code pitfalls. 3x |
| Initial requirements for a no code platform? | The main requirements for no code platforms are their price and their flexibility in terms of features and design. | Benefits during growth | - Speed: - 6c - It is scalable - 6x - Customization - 4x - Funding, access to commmunity and team / founders - 3x - Affordablity / Cheap (Sharetribe) - 2x |
| Which no code stack / platform did you use? | What's notable is, 4 projects didn't use their main no code tool to build their landing page, but instead choose more powerful website builders like Webflow or Wordpress. Most of them also integrate Google Analytics (and others) and sometimes Stripe for payments. Let's keep in mind, that their first version with which they launched. | What would you use | - Use the same tech stack at the start again - all - Bubble: 4x - Perfect for startups in early stage, when sucess it not gurasteed (MVP / First version) - 4x - Cheap / Money saver. 3x |
| What was/is your long-term product strategy? | They either did not have a strategy and first wanted to test demand to go from there (Soft), actively test out features that they wanted to add later (Sharetribe), or purposely build the app scalable and without... (Bubble: eleveezo, thubble)... | | |
| **Reasons for no-code** | | **Challenges / Limitations** | |
| Why did you choose to build and launch your product using the above no code platform? | My findings match well with general perceptions and sales speak of No code platforms: Its fast to build, it's cheap and the platforms do exactly what they need for their use case (marketplace, directory, communities) The importance of an active community was new to me. Especially for a complex tool like Bubble it is very much appreciated. | Challenges for MVP | Softr, Sharetribe and Bubble surve different user needs and applications. - Softr, the easiest and "least" capable of the three, had no challenges at launch, as setup really is intuitive, easy and requires no learning curve. - Sharetribe, a dedicated no-code marketplace platform, has challenges with customization. Setup thereof is easy but creating a unique user experience for your specifc use case is challenging. - Bubble the most advanced of the three (and int he field), has challenges with the learning curve, which later reflects in product quality and scaling abilities. It is not a limitation of the tool, but rather of the skills of the person using it. The entry barrier still seems quite low, but mastering it seems to be challenging. Making the web app responsive was a |
| Did you other solutions before? Why did you decide against them? | Yes: 6x No: 4 People who tried other no code solutions mentioned that either features were missing or the price was too high. | Challenges during growth | Comparing the three platforms, common patterns emerge for each of them. - Softr has challenges with features that builders expect, cost when using large data sets and how to automate a marketplace without it feeling like patchwork. - Sharetribe has similar challenges in terms of features and quickly reaches customization limitations. Yet, users did find workarounds. |
| Why did you decide against building a custom solution? | None of the people interviewed have a custom background and cannot or only have basic coding skills. If they want to build a software enabled business, they are only left with paying someone else to custom code it, have a developer join their team or go to no code. It is no wonder that speed and money are measured as second largest reason why they didn't choose custom code. Especially the businesses who started on no code, also mentioned that no code enabled them to reduce market risk as they can test their solution early in the market, without spending much resources. If the engineering knowledge is not in the company itself, they have to scope out an MVP before they... | What would you change? | The speed of no code, reduces the need for requirements engineering, research and translating it for software engineers. The more complex the no-code platform is, the more skill is requried of the builder. In turn this influences the amount of mistakes that can be made. No-code might change the way we think about building products and user research because the playing field changed. |
| | | **Rebuilding or switching to custom code** | |
| | | Re-building your product with a custom code solution? If yes, why. If not, why? | - Everyone considered it. - 2 are actively building a new solituion wiht custom code. - 3 want to rebuild it, once they cant grow anymore, proove their business or find a CTO. - 4 dont want to rebuild it, because it is a distraction, not needed or stopped working on the project (2)" |
| | | Did you switch or modify your tech stack after launch? If yes, why? | Stayed with no code: 8x Rebuilding it: - Only after proving their business (180k customers + right CTO) - 11k customers, 25kMRR + finding investors that provide a software team) |

Figure 10: *Findings Summary*

**Summary figures:** Summary figures (Fig. 12-17) are based on the meta-analysis. Only items mentioned at least twice are included (Find detailed figures in the appendix). Some figures (Fig. 13, 14, 15) use different letters and/or are in brackets. Each letter indicates one occurrence in the interview.

X = one occurrence in the interview, no-code related.

P = Product-related (something that could occur with or without the use of no-code)

M = Maker-related (something that relates to the maker's mindset and skills)

() = When mentioned elsewhere in the interview or a single argument is grouped into a meta category.

## 5.4 Quantitative Rating Scale Survey

The questions are based on the interview questions and include LCNC benefits and challenges based on the literature. Figure 11 shows the twenty questions used, organised into early days (left) and later days (right).

| Impact of no-code on digital product development | |
|---|---|
| Research Question 1 (RQ1) | Research Question 2 (RQ2) |
| 1.) Why do software startups build, launch and scale their product on top of a no-code solution? 2.) This includes the decision-making process, requirements and long-term product strategy. | 1.) What benefits and limitations do no-code platforms currently have? 2.) Why (if at all) do software startups choose to transition to a custom solution? |
| **Early days: Zero to one.** | **Later days: One to infinity (Growth).** |
| I / we  could launch faster using no code. | Using no code hindered you during growth. |
| I / we could develop a better product using no code. | Using no code helped you grow faster. |
| The no code platform or no code tech stack is affordable. | Using no code helped you to keep up with the rising number of users. |
| I / we could iterate more quickly and easily using no code. | Helped you to move fast during growth. |
| I / we could better validate the problem. | Helped you to test and validate new features easily |
| I / we could better validate the solution. | Helped you to develop new features or products quickly. |
| I / we could validate the value proposition. | Allowed you to implement new features or products the way you wanted. |
| I / we could better validate the business model / revenue streams. | Helped you to save costs. |
| I / we could better validate the marketing channels. | Helped you to stay competitive |
| I / we could attract more customers? | The no code platform / stack was worth the money |

Figure 11:

*Rating Scale Survey Questions*

A numerical rating scale from one (fully disagree) to ten (fully agree) in combination with statements was used. Responses were collected and analysed using Google Forms (docs.google.com). Respondents filled out the survey immediately after the interview.

# 6 Study Results

## 6.1 Background and Context

| Platform | | About | Time to market (TTM) Team size Commitment | Users/Traction | Revenue |
|---|---|---|---|---|---|
| Softr | Interiorly | Marketing info product. Launched August 2021. | TTM: 3 days + 6 month data agregation. Team size: 1 + freelancers. Commitment: Side Hustle | - 270 free users<br>- Monthy subscriptions (39$): 17 (for 5 months)<br>- Switched to one time access for 39€ | Total earnings:4000€ Total money spent: 1200€ |
| | 100 Days of nocode | Online community and bootcamp to learn no-code Launched March 2020 | TTM: 2 days (no plan to turn into business) Team size: 3 Commitment: Full-time | - 200 signups on launch.<br>- 6k email subscribers | - 250-300 paid subscribers (8$/month)<br>- 30 people per bootcamp (750$/person) (4 bootcamps completed) |
| | AIDE Hulp | Marketplace to offer and find help for victims of floods in July. Launched July 2021 | TTM: 2 hours Team size: 2 Commitment: Side Hustle | - 270k page views on first days.<br>- 3k listings<br>- 15k views following weeks.<br>- 6k views the following month | - Did not charge money.<br>- Ran a crowdfunding campaign via external platform. |
| | Climate CollabEU | European climate platform and community for anyone interested. (Investors, Employees, Founders, Newbies) Launched Dez. 2021 | TTM: 2 days + 6 month of data collection Team size: 3 Commitment: Full-time | - 8k jobs on platform (1k new jobs each week)<br>- 3k climate startups (that maker mostly added)<br>- 4k investors (that maker mostly added)<br>- 600 website signups, ~70% joined community<br>- 15k email subscribers, ~ 100 new / week.<br>- 500 page visits from SEO. | - Started experimenting with paid job postings at time of interview. |
| Bubble | Connecto lingo | Arabic/Hebrew Language learning platform Launched: 2018 | TTM: 6 months (next to CEO full time job) Team size: 9 Commitment: Full-time | - 30k users<br>- 10-15% are paying | |
| | Locumify | Job platform for doctors to findlocum work. Launched new ver.: Mid 2019 | TTM: 6 months (next to existing business) Team size: 9 Commitment: Full-time | - 20 recruitment agencies that cover 90% of countries locum jobs.<br>- 100k jobs added to platform<br>- 1500 Doctors | - Self-sustaining<br>- Received 1 Million funding 2 years ago.<br>- One of the most successful companies from Bubble in terms of funding. |
| | Talent Tap | Job platform based on talents next career step. Launched: Januar 2021 | TTM: 1.5 months. Team size: 2. Commitment: Full-time | - 11k talent base<br>- 80 companies<br>- 8 companies subscription, rest sucess bonus | - 10k MRR<br>- 100k esarned this year.<br>- 25k made past month (Highest ever) |
| Share-tribe | Bike2 Loop | Marketplace for second hand bikes and digital store front for bike shops. Launched: Feb. 2021 | TTM: 2 weeks (for basic MVP) 2 months to launch. Team size: 2 Commitment: Side Hustle | - 360k page views since launch<br>- 5000-3000 page views / day<br>- Over 1k listings<br>- 5k users | |
| | Car Mingle | Peer to peer car sharing platform. Launched: 2019 | TTM: 3 months for platform, 9 months for legal. Team size: 50 Commitment: Full-time | - 180k customers<br>- One of the largest customers of Sharetribe | - Aquired a hardware startup.<br>- Received 4.2 million funding. |
| | Paperound | Marketplace to find students for desktop jobs. Launched: January 2021 | TTM: 1 month. Team size: 4 + freelancers Commitment: Full-time | - 50k in student earnings<br>- 500 completed projects<br>- 99.5% sucess rate.<br>- ~ 400 taskers | - Received 150k investment after 3 months. |

Figure 12: *Background and context*

**Profile of the makers:** Most respondents are in their thirties. All of them have a marketing or business background. One is a trained doctor.

They learned about the no-code movement generally in three ways: (1.) while working at a company and trying to improve processes; (2.) accidentally stumbled upon an article or video; (3.) actively researched cheap and quick alternatives to custom-code.

All makers knew the no-code platforms Bubble and Zapier, and most also knew Airtable and Stripe.

The makers worked in small teams of one to four people or established teams of nine to 50 full-time people.

**Category:** Six of the makers run marketplaces, two run a community, and the remaining two an info product and language learning app. This informs the features and requirements found in this research.

**Traction:** All of the projects have significant traction in terms of users, engagement, funding, and or revenue—especially Bubble and Sharetribe Apps. Softr apps received high engagement (page views), but monetisation was less developed (Figure 12). In contrast, all Bubble and Sharetribe startups (except Bike2Loop) are profitable. Notable is Car Mingle, which is one of the largest no-code startups on Sharetribe (180,000 customers) and influenced the Sharetribe product roadmap (p. 10). Locumify has "been one of the most successful companies on Bubble in raising money and being profitable" (p. 9).

**Time to market:** Most products were launched within a few days to a few months (Figure 12). Notably, projects built on Softr.io are the fastest in building the website, but data collection was the most time-consuming part (ClimateCollabEU, p. 2; Interiorly, p. 5). Bubble startups took about six months, even though the makers worked on it next to their existing business. If they had worked full-time, they could have developed it in about two months (Locumify, p.4, Connectolingo, p. 6).
Sharetribe makers reached a time to market of about 1-3 months. In Car Mingles's case, a team of three experienced Sharetribe developers worked full-time for three months on the first version costing $20.000 (Car Mingle, pp. 4).

## 6.2 RQ1: Why do software startups build, launch and scale their product on top of a no-code solution?

### 6.2.1 Initial Product and No-code Platform Requirements

| Requirements | Softr.io | | | | Bubble.io | | | Sharetribe.io | | | Total |
|---|---|---|---|---|---|---|---|---|---|---|---|
| | Interiorly | 100Daysofn... | Aide Hulp | ClimateCollab | Connectolingo | Locumify | Talent Tap | Bike2Loop | Car Mingle | Paperound | |
| **Initial Product requirements** | | | | | | | | | | | |
| Browse Listings (Search, Categories, Sort) | x | | x | x | x | x | x | x | x | x | 9 |
| Filter | x | | x | x | | x | x | x | x | x | 8 |
| Create listings | | x | x | x | | x | | x | x | | 6 |
| Login, Signup and user accounts (with info) | | | | x | x | x | x | | | | 5 |
| Payments | x | | | | | x | | x | x | x | 5 |
| Detailed Page | | | | x | | | x | | x | | 3 |
| Mobile and Desktop friendly | | | | | x | x | | | | | 2 |
| Messaging | | | | | | x | | x | | | 2 |
| **Initial no-code platform requirements** | | | | | | | | | | | |
| Subscription /Payments | x | | | x | x | | | x | x | | 5 |
| Something quick and get up fast. (Easy to fugure things out with) | | | x | x | | x | | | | x | 4 |
| Cost save / cheap: | | | | | | x | | x | | x | 3 |
| Flexibility: Feature rich, freedom, no (perceived limitations) | | | | | x | x | x | | | | 3 |
| Sort, filter, Visualise, organise content | x | | x | | | | | | x | | 3 |
| Native Integrations (Stripe and EMails) | x | x | | x | | | | | | | 3 |
| Flexible Backend / Database (Add logic) | | | x | x | x | | x | | | | 3 |
| Ability to build a fully functional marketplace add listings, and search these listings,pay for it. | | | | | | | x | x | x | x | 3 |
| Decent learning curve: Detailed (Buble dev) course (from Zeroqode) | | | | | x | | x | | | | 2 |
| Payments handled via platform | | | | | | | | x | | x | 2 |
| Community or Forum to ask questions to and find answers | | | | x | x | | | | | | 2 |
| User manage their own content (Login / Accounts) | x | | | | x | | | | | | 2 |
| **Primary Tech Stack** | Softr + Airtable | Carrd + Mailerite | Softr, Airtable, (MVP) Later: | Softr, Airtable, Actito (Mail) | Bubble, Airtable, Phantom Buster | Bubble (App), Articulate Storyline | Bubble (App), Webflow (Website), | Sharetribe .io, Paypal, Stripe, G | Sharetribe | Sharetribe (Marketpl ace), | |
| **Product Strategy** | Yes, more | No, do small | No, timing | No, did lean | Yes, run two | Yes, use | Yes, did lean | Yes, monetise | Yes, add | No | Yes (6x) |

Figure 13: *Initial Requirements*

This section explores what features makers looked at before starting to build. It turns out that many of them will be mentioned as benefits later; however, this describes their initial need.

**Product requirements:** The top product requirements are basic marketplace features such as creating and browsing listings, accounts and payments. Detailed pages, mobile/desktop friendliness and messaging were important for some (Fig. 13). Given that six out of ten are marketplaces, this result is not surprising given the type of no-code startups in this research.

**No-code platform requirements:** The no-code requirements here are less homogeneous. The top requirements are built-in payments/subscription handling, a platform to build something fast, easy to learn and cheap. All Bubble users mentioned flexibility and freedom to build anything, and most Softr makers mentioned the importance of native integrations like Stripe and analytics.

## 6.2.2 Reasons for No-code

| Reasons | Softr.io | | | | Bubble.io | | | Sharetribe.io | | | Total |
|---|---|---|---|---|---|---|---|---|---|---|---|
| | Interiorly | 100Daysofn | Aide Hutp | ClimateColab | Connectolingo | Locumify | Talent Tap | Bike2Loop | Car Mingle | Paperound | |
| **Why did you choose this nocode stack?** | | | | | | | | | | | |
| Feature set: Perfect feature set to what they needed. | x | | x | | x | | | x | x | | 6 |
| Speed: very easy to quickly build something | | x | x | | x | | x | | | x | 5 |
| Cheap. Good to test idea, value for money | | | x | | | | | x | | x | 4 |
| Existing community + Forum with experts to ask questions & get answers | | | | x | x | x | | | x | | 4 |
| Social proof: People built something similar or recommended it. | | | | x | | | x | | x | | 3 |
| Customization and design flexibility | x | | | x | | | x | x | | | 2 |
| Well funded startup / Features on the roadmap / only getting better | x | | x | | | | | | | | 2 |
| (Native) Integrations (Stripe e.g.) | x | | | | x | | | | | | 2 |
| Familiarity. Had used it before | | | x | | | | | | | x | 2 |
| **Why did you decide against custom code.** | | | | | | | | | | | |
| Doesn't know how to code | (x) | x | x | x | x | (x) | x | x | (x) | (x) | 6 (10) |
| Speed: No-code is much faster to build (Dev team too slow) | | | x | x | x | x | | | x | | 5 |
| Cheaper (Dev team is / was very expensive) | | | | x | x | x | | | x | | 5 |
| High risk with getting requirement right. | | | | | x | | | | x | | 3 |
| Able to build on their own, without having to wait for someone else. | | | | x | | x | | | x | | 3 |
| **Did you try other solutions?** | Yes, alternatives would need workarounds which increase complexity. | No, did know Softr existed at the time | No, worked with this stack before and knew it perfect for this use case. | No, trusted recommendation s of people who built sth. Similar existing courses. | Yes, MS Silverlight and .NET before. Could use existing courses. | Yes, had dev team before. They couldn't build a good product fast. | No, tried the free version and realised in can do everything he | Yes, explored custom code agency starting at 10k€, out of budget | Yes, custom built a solution in parallel to find out they needed sth. faster. | Yes, explored marketplace software's and tried custom code before. | Yes- 6x No- 4x |

Figure 14: *Reasons for no-code*

**Reasons for the no-code stack:** The top reason for the chosen no-code stack was the feature set (Fig. 14). It shows that the platforms solve the different builders' needs. Others among the top are development speed and price, as well as the importance of an active community where the makers find expert answers. This ties into available examples of similar no-code products that serve as inspiration (social proof). "If they can do it with a sauna, you can do it with a car as well" (Car Mingle, p. 2).

**Reasons against custom-code:** The reverse question of "why they decide against custom-code" reveals an exciting insight. None of the makers has a technical background. A few have basic knowledge but not enough to build complex applications. Makers with little to no software development knowledge have three options. (1) pay a developer to custom-code it, (2) have a developer join their team, or (3) use a no-code platform. Scoping the correct product requirements is challenging, even with proper user research. No wonder cost, speed and the fear of developing the wrong product are the biggest downsides from their point of view.

"If you want to build something [custom], you need to specify very much what you want. If you need to work with someone or a group of people, you need to trust that they do the right thing. I'm not an engineer, so it becomes much harder" (Car Mingle, p. 3).

**Other solutions:** Six of the makers tried other solutions before. Those who tried custom-code mentioned price (too expensive) and development speed (too slow) as reasons against it. Connectolingo and Locumify, for example, ran on a custom-code solution before switching to Bubble. Bubble was faster and cheaper. "We had this big development [4] team that we were spending \$1000s [...] of pounds on every month. [...] We just could not rely on our development team to build it because it was too slow" (Locumify, pp. 4).

Makers who tried other no-code solutions mentioned the lack of features (such as integrations) and bad value for money as reasons against it (Fig. 13).

The four makers who did not try other solutions were confident that their chosen platform would suit their needs.

## 6.3 RQ2: What benefits and limitations do no-code platforms currently have, and why (if at all) do software startups choose to transition to a custom solution?

This question looks at the benefits and limitations of no-code across the entire startup journey. Tailoring to the second hypothesis, that with growth, no-code startups need to switch to custom software.

### 6.3.1 Benefits

| Benefits | Softr.io | | | | Bubble.io | | | Sharetribe.io | | | Total |
|---|---|---|---|---|---|---|---|---|---|---|---|
| | Interiorty | 100Daysofn. | Aide Hulp | ClimateCollab | Connectingo | Locumify | Talent Tap | Bike2Loop | Car Mingle | Paperound | |
| **Benefits for MVP** | | | | | | | | | | | |
| Speed | | x | x | | x | x | x | x | x | x | 8 |
| Cost | | | | | | x | | x | x | x | 4 |
| Learning curve, community and forum | | | | | x | x | x | | | | 3 |
| Quick product validation and user testing | | | | | | x | x | | x | | 3 |
| Easy to use | | x | | | | | | x | | | 2 |
| predefiend UX, blocks and templates (positive | | x | x | | | | | | | | 2 |
| Removes complexity (Visual Front + backend, no bugs or common coding pitfalls) | | | x | | | x | | | | | 2 |
| Fimiliarity / She knew the tool | | | x | | | | | | | | 1 |
| Flexibility | | | | | x | | | | | | 1 |
| No "translation" of requirements needed. | | | | | | x | | | | | 1 |
| **Benefits during growth** | | | | | | | | | | | |
| Speed (Test new features / make changes, positive contraint | x | x | x | | | x | x | | x | x | 7 |
| Scalable | x | | x | | x | (x) | x | | | | 5 |
| Platform trajectory (funding, community, contact to team) | | x | | x | | | | x | | | 4 |
| Stability /reliability (removes common code problems) | | | | | x | x | | | x | | 3 |
| Customisation | x | | | | | | | | | | 2 |
| Affordable, Cheap | | | | | | | | x | | x | 2 |
| adding features is easy | x | | | | | | x | | | | 2 |
| Easy to use | | x | | | | x | | | | | 2 |
| **What would you keep?** | | | | | | | | | | | |
| Start with the same tech stack again | x | x | x | x | x | x | x | x | x | x | All |
| Speed (positive restraints) | | x | x | | | x | | | x | x | 6 |
| Being lean: Perfect for startups when sucess is not guranteed | | | | | | x | x | x | | | 3 |
| Cheap / Money saver | | | | | | x | | | x | x | 3 |
| Integrations | x | | | x | | | | | | | 2 |

Figure 15: *Benefits of No-code*

**Speed:** The top benefit across the entire life cycle is development and iteration speed. Time to market of a few days (Softr) to a few months (Bubble and Sharetribe) shows that (Fig. 15). The peer-to-peer support marketplace Aide Hulp is a prime example. "The 14th [July 2021] is when the floodings happened. On the 15th, we got the idea to build something, and we launched in the afternoon" (Aide Hulp. p. 10). They were quicker than official instances and even faster than first aid organisations like the Red Cross. In the example of Paperound, building on top of an existing solution saved them about one year of development time (p. 4). Softr's ready-to-use blocks speed up the building process and provide "positive constraints". "I can't design all day there, but it gets the maker "80% there" (100 Days of no-code, p. 9, Aide Hulp, p. 5).

Even in the later stage, speed is mentioned as a benefit. Makers can test new features quickly with real users and data using the live product. (Talent Tap, p. 4) "The speed is disruptive. [...] It is some alien technology" (Talent Tap, p. 7).

**Cost savings:** Cost was primarily mentioned by Bubble and Sharetribe startups. Locumify freed up an extra year of runway by switching from code to Bubble (p. 6), and Paperound got its first users on the free trial of Sharetribe (Paperound, p. 2).

**Scalability:** Against the common misconception that no-code cannot scale (Alamin et al., 2021), scalability was mentioned as the second biggest benefit during growth by all Bubble makers (Fig. 15). However, the ability to scale is highly dependent on how the app is built which requires knowledge about basic software development practices from the maker (Fig. 16). If done wrong makers end up with technical debt that needs to be solved later (Talent Tap, p. 5).

**Platform trajectory and community:** The fact that the no-code platform recently received funding encourages makers to choose this platform over alternatives. They feel it is future-proof and only getting better (Connectolingo, p. 8). Close access to the founding team, experts and learning materials, and an active community provide peace of mind (ClimateCollabEU, p.4).

**What to keep:** Potentially, the biggest indicator for the validity of the benefits is the fact that all the makers would start with the same tech stack again (Fig. 15). Being lean and learning from the market quickly without high financial commitments are worth the potential limitations that might arise during growth.

## 6.3.2 Challenges and Limitations

| Challenges and limitations | Softr.io | | | | Bubble.io | | | Sharetribe.io | | | Total |
|---|---|---|---|---|---|---|---|---|---|---|---|
| | Interiorly | 100Daysofn... | Aide Hulp | ClimateCollab | Connectolingo | Locumify | Talent Tap | Bike2Loop | Car Mingle | Paperound | |
| **Limitations & Challenges MVP** | | | | | | | | | | | |
| No limitations | x | x | x | x | | | | x | | | 5 |
| Skill of the maker (learning it/make it scalable) | | | | | x | x | x | | | | 3 |
| Customising out of the box platform | | | | | | | | x | x | x | 3 |
| Mobile responsive engine (old Bubble Engine) | | | | | x | | x | | | | 2 |
| Integrating many different services | | | | | x | | | | x | | 2 |
| UX (slow loading pages, customisation limits) | | | | | | x | | | | x | 2 |
| Lack of tutorials & ressouces (when they started) | | | | | | x | x | | | | 2 |
| Integrating payments in their country | | | | | | x | | | x | | 2 |
| **Limitations & Challenges Growth** | | | | | | | | | | | |
| Skill of the maker (Risk of building unscalable apps) | | | | | x | x | x | | | | 3 |
| Long workflows | | | | | x | | x | | | | 2 |
| Limited insights | | | | | | | | x | | x | 2 |
| Multiple payment methods | | | | | | | | x | x | | 2 |
| Design limitations / Customisation | | | | x | | | | | x | | 2 |
| How much should they automate (PQ) | | P | P | | | | | | | | 2 |
| **What would you change?** | | | | | | | | | | | |
| Maker activities | | | (M) | | | (M) | (M) | (M) | (M) | (M) | 6 |
| Product decisions (in hinsight) | (P) | | | (P) | P | (P) | | | | | 4 |
| Plan / know Data structure from start | | | | M | M | M | M | | | | 4 |
| Tools wise, I wouldnt change allot | x | x | | x | x | | | | | | 4 |
| No product requirements planning | | | M | | | | M | | | | 2 |
| Team (find good co-founder, CTO) | | | | | | | | M | M | | 2 |

Figure 16: *Challenges and Limitations*

The challenges and limitations turn out to be very platform specific as Softr, Sharetribe, and Bubble serve different user needs and use cases. Generally, makers mentioned more product and maker-related challenges that would apply to no-code and custom-code products (Fig 16).

**Softr** makers had no challenges at launch. However, they faced project-specific challenges after launch (Detailed figure in Appendix 11.5.2). For example, multi-language support, conditional sub-filter, advanced SEO, integrated payment portal (now available), cost when using large data sets and automating a marketplace without feeling like patchwork. After experiencing growth, Aide Hulp "magically worked" but "felt glued together" (p.9). "Especially when you want to bootstrap [and you have over 10,000 data entries per database] is quite a bit of money at the beginning" (ClimateCollabEU, p. 5).

**Bubble**, the most versatile of them, has challenges with the learning curve. The entry barrier still seems relatively low, but building a scalable Bubble App seems to be challenging without basic (software) knowledge. It is not a tool limitation but rather the maker's skills. All the Bubble makers mentioned the importance of understanding database structures and front/backend logic as scalability factors. Most of them did face some form of technical debt, that they were able to resolve as their skill improved. This is one reason why they value access to an active community. Bubble-specific challenges include making the web app responsive, which the new responsive engine addressed in June 2022 (*Bubble.io*, n.d.).

**Sharetribe** is easy to set up but faces challenges with customisation, automation and providing insights into the platform (depending on the Sharetribe plan). Creating a unique user experience for the specific use case is still time-consuming and challenging (still less than custom-code); however, makers found creative workarounds. "We're trying to run a business on level three, using the architecture of level one. But, we saved a lot of money and time getting from level zero to level one" (Paperound, p. 5).

**What to change:** A lot of what makers would change are product decisions (P) or maker-specific activities (M) but would not change a lot regarding their no-code tech stack (Fig. 15, 16). Product decisions they would change include better user research upfront, more feedback collection after the launch of the MVP, testing monetisation earlier or better planning of product requirements before building. Maker-related changes include team questions like finding a CTO earlier or better documentation (Appendix 11.5.2).

### 6.3.3 Rebuild or Switch to Custom-code

This section looks at when, why and if no-code startups switch to custom-code.

| Transition to custom code | Softr.io | | | | Bubble.io | | | Sharetribe.io | | | Total |
|---|---|---|---|---|---|---|---|---|---|---|---|
| | Interiorly | 100Daysofnoc | AIDE Hulp | ClimateCollab | Connectolingo | Locumify | Talent Tap | Bike3Loop. | CarMingle | Paper Round | |
| # Features added/changed after launch | 0 | 8 | 4 | 4 | 4 (everything) | 3 | ? | 5 | 2 | 3 | |
| Switched or modified tech stack? | No | Yes | No | No | No | No | Yes | No | Yes | No | No: 7\| Yes: 3 |
| Rebuilding it in custom code? | No | No | No | No | No | No | | No | | No | No: 8\| Yes: 2 |
| Reason why? | No need / product market fit | No need, if at all rebuilding it in Bubble. | Considered it. Stopped working on project. | Considering it. #1 Prio: finance, When CTO maybe. | No need, everything works. | No need | Found investors that bring software team. | Considered it. Once business is validated, build it custom. | Improve speed, social logins, multiple payment methods. | Considered it. As long as they can grow, no need | No need: 4 Considered: 4 Rebuilding: 2 |

Figure 17: *Transition to Custom-code*

Only two of the ten startups interviewed actively rebuild their product with a custom-code solution. Four actively considered rebuilding it but decided against it for the time being. Two startups (Connectolingo and Locumify) used custom-code before. They switched to no-code (Bubble) (Fig. 17). Almost all added features or changed functionality after launch without encountering major limitations, which speaks to the no-code benefits. More details can be found in the one-page summary in the appendix.

**From code to no-code:** Connectolingo and Locumify makers built their Bubble app over six months while maintaining their existing business (p. 6, p. 4). In the case of Connectolingo, their custom-code environment .NET was discontinued, and they needed an alternative. Money "was an issue back then" (p. 7), and Bubble integrated into their existing courseware, which is why they chose it. With the maker's software experience, he was confident Bubble could scale if he built it correctly.

The story of Locumify is similar. They realised that their custom-code product was not working, so they started experimenting with a new twist to the product and built it on Bubble (Fig. 12). They realised that one person could build the entire Bubble app "maybe 20 or 30 times faster, than [their] development team" (p 1.). That person was the domain expert, product manager, and developer all in one, which removed the need to "translate" requirements (p. 6).

Notably, they either had a basic software background or learned Bubble through trial and error to arrive at the skill to build a scalable app. They do not intend to go back.

**No intent to switch:** The maker of 100 days of no-code thinks that "building is a distraction" (p. 8). His main value is building a strong community, not the best technology. The maker of Interiorly still struggled to find product market fit; therefore, he had no intention to switch.

**From no-code to code:** Car Mingle is a fascinating example of switching to a custom-code solution. They wanted to build a custom solution at 100,000 users (an arbitrary number) but did not find a suitable CTO to manage that process. Only when they found a CTO, at 180,000 customers, they started (p. 7). They outgrew Sharetribe and hit limits that compromised customer experience. They needed features such as social logins, multiple payment options (available in Singapur) or live location tracking of cars - features that Sharetribe would not add anytime soon.

Talent Tap started fundraising for their no-code job platform. Eventually, they found investors with access to an existing software development team that offered them funding if they would rebuild it in custom-code. They took that offer. Also, they started to hit growth barriers. For example, providing GDPR compliant APIs to applicant tracking software like Personio (personio.com) would have been difficult to achieve on Bubble. One reason they had to decline some German brands

**Considering it:** The makers of ClimateCollabEU and Paperound seriously considered switching to custom code. For ClimateCollabEU, the priority was finding revenue streams first instead of improving the platform. Still, they felt limited by the predefined blocks and lack of customisation, which compromised user experience - the same features that made building an MVP blazing fast.

Paperound slowly starts to experience the limitations of Sharetribe, similar to Car Mingle. They still run on the basic plan (Sharetribe Go) but built scrapers and connected them to AWS buckets to automate processes and retrieve insights that Sharetribe does not provide (on this plan). "The original plan was to use Sharetribe, raise investment, and build it properly. [...]. If we can still grow with the basic tech, [...] I'm just happy to keep costs down. [...]. [When] we struggle to improve conversion metrics, then it is time to rebuild [...]" (Paperound, p. 5).

## 6.4 Summary

Both of the hypotheses could be verified based on the results. However, the answers have more depth than anticipated.

**Reasons for no-code and benefits:** The biggest drivers for makers to build on no-code are their lack of software engineering knowledge, unrivalled speed and cost savings. No-code platforms precisely solve the maker's initial needs and reduce risks, making them their first choice, even when they can custom-develop software.

**Challenges and Limitations:** Challenges and limitations are platform and startup specific. Softr did not face challenges when building the MVP but later lacked some features and customisation. Bubble's biggest limitation is the makers' skill and their learning curve to build scalable apps. When this is solved, Bubble is the most versatile of the three. Sharetribe faces challenges with customisation and limited insights after launch (depending on the plan).

**Switch to custom-code:** Only two startups interviewed are actively rebuilding their product in custom-code. The big difference is that they do so with reduced market risk. They received funding, have revenue or customers. Growth metrics seem to decide whether to rebuild it in custom-code. They consider custom-code once they cannot grow users or retention metrics drop due to product limitations.

## 6.5 Quantitative Survey

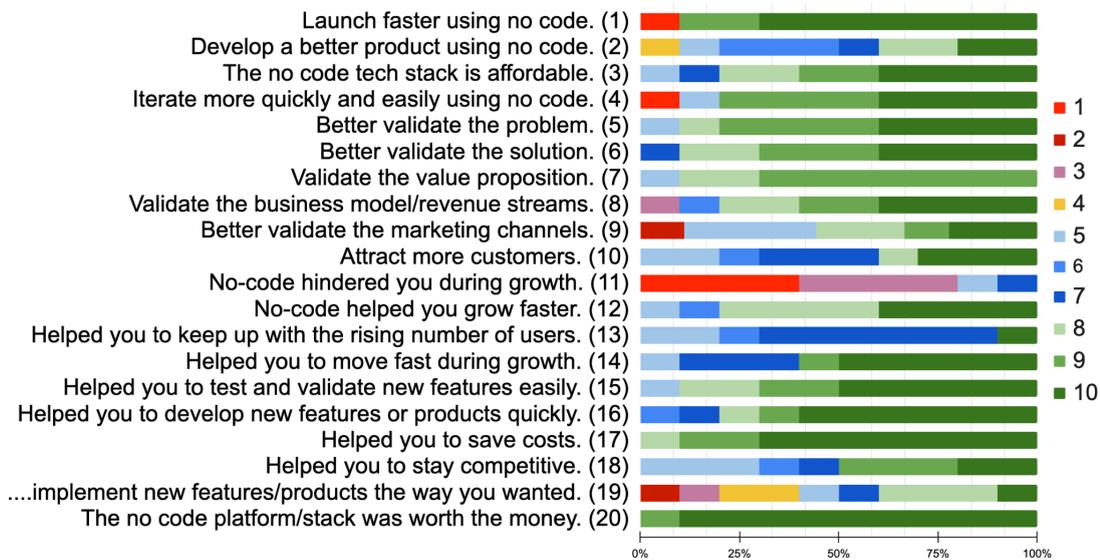

*Figure*

*18: Quantitative survey results (1: No, fully disagree; 10: Yes, fully agree)*

The quantitative questionnaire results (Fig. 18) largely back and enforce the interview findings. The majority shows a high agreement to the statements (rating 8-10: yes, fully agree), with a few leaning towards the mid ranges (questions: 2, 9, 10, 13, 14, 18) and no clear consensus for questions 2 and 19.

The limitations around the makers' skill, scalability, and customisation are endorsed (questions: 13,19), as well as the various benefits like a leaner beginning, speed and cost savings (questions: 1, 3-8, 17, 20). There is no clear consensus around no-code's ability to build a better product across the entire life cycle.

# 7 Discussion and Limitations of Research

This section explores the findings of this study within the context of available literature and the connection between digital product development and established product management practices. The results of this study match the identified no-code benefits and some of the limitations in LCNC literature. Additionally, it presents new insights between no-code and product development.

## 7.1 Why build on No-code? - Benefits of No-Code

The identified benefits and why makers choose no-code are interlinked.

**Driver for no-code:** The biggest driver for no-code is that none of the makers has a professional software background (Fig. 12). No-code removes traditional barriers between the maker and a software product. Barriers like cost to hire developers, long development times and high risk due to wrong scoping translate into no-code benefits. For the first time, non-technical makers can build without engineers and have "control over their product" (Bubble Group, Inc., 2020). Additionally, no-code platforms have gotten more diverse than ever, tailoring an increasing number of use cases (Palios, 2022). When it comes to choosing the right no-code platform, the features, community, learning resources available, and funding of the platform seem to be important to the makers (Fig. 13).

**Benefits:** The two biggest benefits of LCNC platforms are faster build time (speed) (Alsaadi et al., 2021; Bubble Group, Inc., 2020; Luo et al., 2021) and reduced cost (Bubble Group, Inc., 2020). Although "ease of use" is mentioned in LCNC literature (Alsaadi et al., 2021; Bubble Group, Inc., 2020; Luo et al., 2021) and influences speed, it did not stand out in this study. In contrast, Softr and Bubble makers mentioned scalability as one benefit which does not match LCNC literature. Only people who did not use LCNC see scalability as a limitation (Alsaadi et al., 2021).

As discussed below, the benefits seem to be a symptom of less tangible observations. Makers get market feedback earlier, combine multiple roles and build technology-enabled products.

## 7.2 Why transition to custom-code? - Challenges and limitations of No-code

The available low-code literature generally agrees on two challenges and limitations that also match this study's finding. No-code platforms have limitations, but they depend on the type of product, the no-code platform and the maker's skill.

First, customisation (Alamin et al., 2021; Alsaadi et al., 2021), the extensibility of existing features (Sahay et al., 2020) and the lack of features have been challenges for Softr and Sharetribe makers (Fig. 16). Although both platforms rely on pre-built blocks and templates, they offer to extend functionality (custom-code blocks (Softr) and Sharetribe Flex). However, software knowledge is required.

Second, the lack of knowledge about the platforms (Alsaadi et al., 2021) and the high learning curve (Sahay et al., 2020) match the challenges that Bubble makers experienced (Fig. 16). It is no surprise that the needed knowledge increases with increasing complexity. The entry barrier is low, but building scalable apps requires knowledge. "You can build a shitty site that won't move with and without no-code if you don't understand how the machine works" (Connectolingo, p. 1). "It's not about the language that you use, it's about the logic that you build, and [it] is the same whether you're writing it in JavaScript or Bubble" (Locumify, pp. 6).

Although the makers wished to learn certain software development principles like relational databases sooner (Fig. 16), it is unclear whether this is because of a lack of resources (Alamin et al., 2021).

Limitations like security concerns (Oltrogge et al., 2018) or vendor lock-in (Sahay et al., 2020) have not been mentioned.

## 7.3 The Impact of No-code Platforms on Digital Product Development

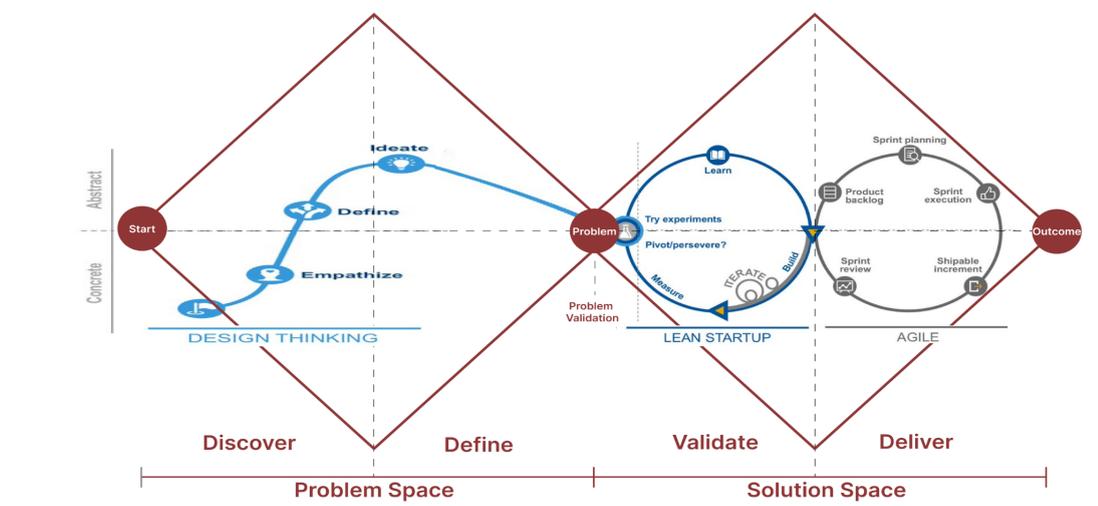

Figure 19: *Design Thinking, Lean Startup, Agile, and the Double Diamond.*
*Adopted from Gartner (Blosch et al., 2016)*

Based on this research, there may be a close relationship between established product development processes, startups and no-code. Product development is often messy, and frameworks or methodologies only sometimes capture this reality. Figure 19 shows an adoption of Gartner's attempt (Blosch et al., 2016) to map out three popular product development approaches: design thinking (with the addition of the double diamond), the lean startup method and agile.

Usually, makers try to find and validate a user problem or need (problem space) and then explore a range of solutions by iterating to settle on the most promising solution.

The goal is to "reduce uncertainty in your business" (Bland & Osterwalder, 2019, p. 82) by "learning faster than anyone else" (Ries, 2011, p. 112) and testing the riskiest assumptions first. According to Ries (2011), they are the value hypothesis, which tests whether a product or service delivers value to customers once they are using it and the growth hypothesis, which tests how new customers will discover a product or service (p. 61).

Ideally, the maker knows that the solution will work, and this certainty is especially important when a lot of resources (time, money, human capital) are needed to build the solution. The top reasons startups fail (Fig. 20) can often be linked to a failed growth and value hypothesis and the benefits of no-code found in this research.

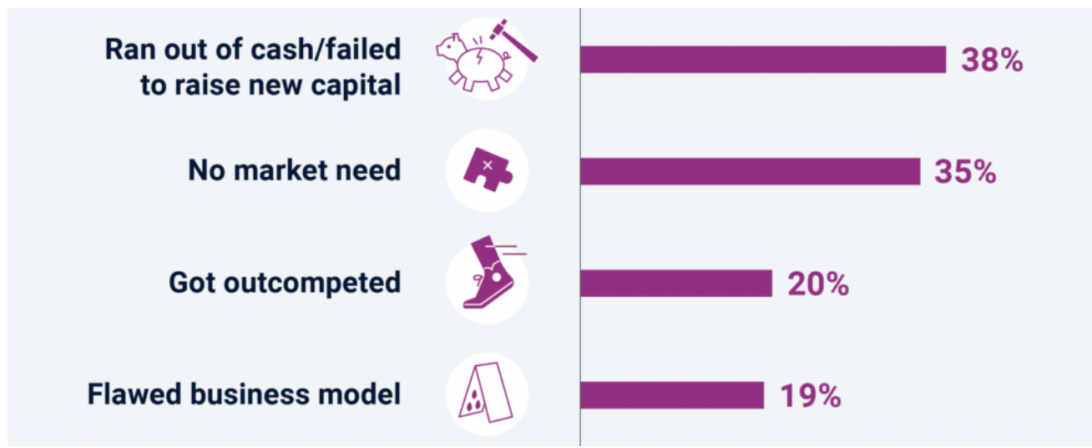

Figure 20: *Top Reasons Startups Fail (CBInsights, 2021)*

**Time to market, feedback cycles and MVPs:** According to Cagan (2017), an MVP should be a prototype and not a product because it leads to a waste of time and money (p. 30). What if no-code could change that?

A shorter time to market means getting feedback earlier and learning faster. "One major benefit about no-code is for proving your MVP. I spent max $500 on the whole business before I raised investment to launch it" (Paperound, p. 2).

If a functional product can be built within a few hours or days (Fig. 12), makers could flip the script, launch a product (based on assumptions) and learn from actual users what they really want. "I can just build it in half an hour. Why don't I just do it instead of talking? If you add actual user research to the process, it almost feels like a delay" (Aide Hulp, p. 12).

Limitations like design flexibility and customisation (Fig. 16) might even be beneficial at the beginning (100 days of no-code, p. 9). "If you're solving a big enough problem, then it doesn't really matter what it looks like. Whereas if you have the most beautiful app [...], you don't know whether you're solving a genuine problem (Locumify, p. 5). Makers should worry about making the product scalable and beautiful once they have traction and improving the product using real user feedback. This way, they reduce the risk of building the wrong product (CBInsights, 2021).

**The maker combining multiple roles:** Another unique no-code advantage is the reduced need to translate requirements between people. In agile software development, the product manager identifies requirements based on user research, creates a backlog and writes user stories that the developers and designers will use to build a feature. Communication and translation between people can be lossy and create overhead. However, no-code enables non-technical people to create software. The maker can be the domain expert, product manager, designer and developer in one. "Having a doctor, being able to design and build the products is [really meaningful]. There are hundreds and hundreds of small details that I put into our product to make it what it is for doctors because I know what a doctor is thinking. The developer will never take that initiative. It's enabled experts to become developers" (Locumify, p6).

**Technology-enabled businesses:** Many no-code startups do not set themselves apart through their unique tech but through their unique market positioning, often fueled by the makers' domain expertise. "We're not doing any deep tech, we're not doing AI, we are not doing blockchain" (Locumify, p.3). They are technology-enabled, not innovating the core infrastructure. The underlying technology has become a commodity (marketplace, community, directory, web and mobile apps). They need to and can set themselves apart through unique positioning. "No one else has the domain expertise that we have to be able to build it [...], which is our competitive advantage, as well as the fact that we can build and ship quickly' (Locumify, p.3). "What was going to prove the business was, have I got the service slotted into the market the right way (Paperound, p.5). "I need to focus on community building, rather than no-code building" (100 Days of no-code, p. 8). Because they are technology-enabled businesses run on no-code, they are faster and cheaper than their custom-code competitors.

**More makers and niche products:** No-code enables anyone (e.g. experts) to build software which might lead to more niche products being built. These products might not be venture cases but might be enough to support the maker and its small team. As the development time for software-enabled products decreases, finding users from a specific niche, in other words, the need for marketing, might increase (Nira Team, n.d.).

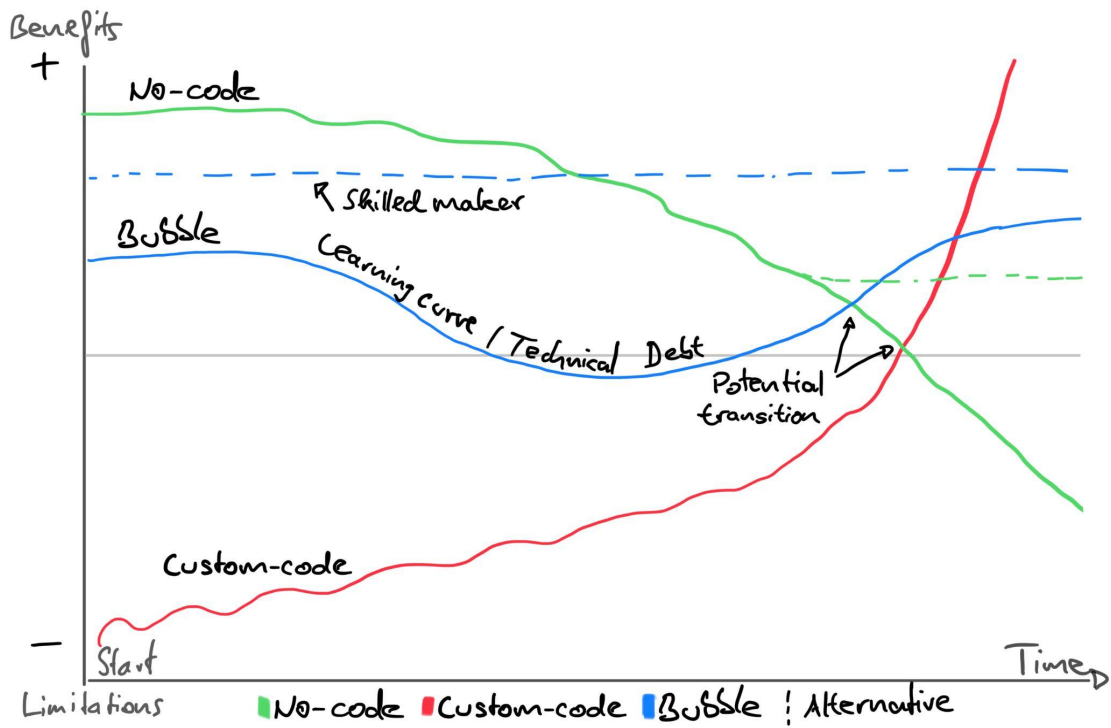

Figure 21: No-code compared to custom-code over a project lifecycle. (Characterises tech-enabled products like marketplaces, directories, and web and mobile apps).

**Key takeaways (Fig. 21):**

**1. Start with no-code:** Makers trying to build a tech-enabled product or business (marketplaces, communities, directories, web and mobile apps) should always try to start with no-code. The benefits (fast time to market, cost savings, maker combining roles) outweigh the limitations in the beginning.

**2. Consider a custom solution once the business is validated:** As time goes by and the project progresses, many critical assumptions should be validated. With a thriving product, limitations might (or might not) become more pressing depending on the no-code platform and product. When product growth slows, transitioning to a custom solution with endless flexibility might be helpful. However, then, the business risk is reduced to a minimum.

All makers would start with no-code again, despite the limitations (Fig. 15).

"Build the product and put it in front of users as fast as you can. This is the only way you can learn – worry about scaling later. For starting something, it's perfect" (Connectolingo, p. 10).

## 7.4 Research Limitations

Three limitations could be identified for this research.

First, the sample size of ten semi-structured interviews might carry low validity. However, the findings are mostly homogenous, especially between the no-code platforms and match existing research, which indicates accuracy.

Second, some interview questions were left out entirely because the interviewee had answered them before. During the analysis, this might have led to occasional double assignments of answers to specific questions. It did not affect the broad implications of this research but could affect specific evaluations of single interview questions.

Third, the selection of the sample might be biased. The no-code platforms featured most samples (Car Mingle, Bike2Loop, Paperound, Locumify Aide Hulp, 100 Days of no-code). Therefore the number of no-code startups that switched to custom-code might be underrepresented in this study. Additionally, the diversity of no-code startups in terms of use cases and users might not cover the full range.

## 7.5 Further Research

This paper is the first to explore the impact of no-code on digital product development in the context of startups and makers.
Three areas for further research might be of interest. The concept of no-code startups and impact on the ecosystem, the role of the maker combining traditional product roles like developer, domain expert and product manager, and the limits of existing and emerging no-code use cases such as AI, Blockchain and IoT.

# 8 Conclusion

This paper aims to answer why makers choose no-code platforms to build, launch and scale a software product, what benefits and limitations no-code has, and why they might transition to custom-developed solutions later. It provides new findings regarding the connection between no-code and established product development.

The research has shown that most makers building with no-code have no technical background. They choose no-code primarily for its faster time to market and cost savings with implications for established product development. Although makers face no-code platform-specific challenges as they grow, they can build a working product and test, validate and grow it with actual users. Additionally, most makers combine multiple traditional roles like software developer, product manager, domain expert, designer or marketeer into one person, which removes much of the "translation" needed in a traditional software team. Therefore, all the makers interviewed would choose no-code to start with again, especially when high uncertainty is involved.

As the field is relatively new, not many no-code startups exist, and selection was influenced by no-code platforms featuring the selected startups, the sample might be biased and not large enough. Further research about the limitations of specific no-code platforms and the maker's role is needed to strengthen the validity of the result.

# 9 Access to Research Data

Access the high-resolution figures [here](here).

Some of the personal information about the makers and startups where anonymised for this public version. However, it is possible to access the original data through a signed NDA. Please contact me at: contact[at]heuschkelsimon.com, to access the following:

- Interview recording
- Interview transcripts
- Interview analysis (excel sheet)
- Rating scale survey results (excel sheet) and form layout
- List of projects and startups analysed

# 11 Appendix

## 11.1 A one-page summary of each interview

### 11.1.1 Interiorly

**[Startup and maker name is anonymised]**

**Maker**: Shrikar (1 person team) | **Platform**: Softr | **Launched**: August 2021 **Type:** Info product | **Time to market:** 3 days (with 6-month data aggregation

**Description:** Interiorly helps interior design business owners get all their marketing content in 1 hour/week. It provides 1000+ templates for all social media and marketing activities. It is a collection of semi-customised templates of pictures, captions, Canva files, and tags for businesses below 3 million dollars.

**Traction:** 270 free users signed up, 17 monthly subscriptions over five months ($700 MRR). Operating expenses: $100/month. Paid about $500 to the freelancers. Spent a total of $1200 and earned around 4000$ at the time of the interview.

**From idea to launch:** The maker (33yo) has a background in marketing and works as a freelance (marketing) consultant. After about three days of brainstorming, he reached out to 200 interior design business owners, from whom he interacted with about 70 and talked to around 30 within two weeks to understand the target audience.

Since this product lives from the content, he hired three part-time freelancers that scraped and created around 1000 pieces and relevant emails for the sales process during the launch. The data collection took around six months.

With all the data in Airtable, he built the first version in Softr in a few hours. He needed accounts, a payment wall and options for the user to search and filter all the content pieces. With all the iterations, he finished the product in one week.

**After launch:** Initially, he offered a monthly subscription (§39) and planned to add more categories and content pieces later if the product succeeded.

At the time, Softr did not offer a payment block, which meant that users had to go to a different site (Stripe) to manage their subscriptions and payment. Users also thought they would get ready-made instead of "semi-customised" content pieces. This led to high customer service volume, and he had to answer more emails than planned. He wanted this to be a low-maintenance side project. Because of this, he switched to a one-time access fee of $39 after five months.

**Keep or change:** He would keep the tech stack with all the native integrations and ask deeper questions during research. Additionally, he would gather more feedback after launch to improve his product based on user feedback.

## 11.1.2 100 Days of No-code (100daysofnocode.com)

**Maker**: Max Haining (3 ppl. team) | **Platform**: Softr | **Launched**: March 2020
**Type:** Community | **Time to market:** 2 days
**Description:** 100 Days of no-code offers three products. A free email course of daily bite-sized lessons to learn no-code, a (slack) community with additional perks for $8 per month and a 30-day boot camp with in-person input for $750 flat.
**Traction:** 200 signups at launch, 6.000 newsletter subscribers, about 300 paid members, and four completed boot camps, with each around 30 people.

**From idea to launch:** The maker (24yo) has a background in business and politics and got inspired about no-code by a Rhyan Hoover (Product Hunt founder) article. He set himself a challenge and wanted to tweet about no-code for 100 days. During that time, he set up a landing page using Carrd.io and connected it with an email to build his first product (the email no-code course) within two days. Although he only had a few hundred followers on Twitter, 200 people signed up when he launched it. It was such a good sign that he committed to pursuing it full-time.

 **After launch:** Since the initial launch, he built out the products he offers today in a team of three. He then launched a paid Slack community using the tipping service Buymeacoffee to manage payments. Since then, he has switched payment providers three times (Gumroad: bad account mgmt, Stripe: low fees and Softr integration, Podia: users can manage their accounts). After around one year, he switched from Carrd to Softr. With Softr, he could add additional perks to paid members and Bootcamp participants. Things like a no-code tool and member directory, past recordings, curated learning content and more.

 **Keep or change:** He would keep Softr as a tool as it allowed him to move fast. It is feature rich and positively constrains him via the pre-built blocks. What annoys him is that the members are split across four separate payment providers. He would not necessarily change it as it worked for him at the time. However, it is with considering: are the products you are picking initially going to serve you long term?

### 11.1.3 Bike2Loop

**[Startup and maker name is anonymised]**

**Maker**: Mark (2 ppl. team)| **Platform**: Sharetribe | **Launched**: Feb. 2021

**Type:** Marketplace | **Time to market:** 2 weeks for MVP; 2 months until launch.

**Description:** Bike2Loop is a marketplace to find new and second-hand bikes and parts. They can be submitted by individual users as well as by bike shops.

**Traction:** 350k page views since launch. 500-3,000 page views/day. 1,000+ listings and over 5,000 users.

**From idea to launch:** The maker (33yo) has a business background and started this project with a friend as a side hustle to solve his itch. In August 2020, he had the idea and started brainstorming ideas. In November, he bought the domain and built the first MVP using Sharetribe in two weeks over Christmas. During January and February, they finetuned functionality and filters and started to acquire the first bike shops. Soon they learned that bike shops wanted to sell their new bikes too. The big launch happened end of February when a local news article heard about it and featured them for free. They received over 40,000 page views in the first days but only had 15 bikes on the platform. The launch was "too early". They also explore hiring agencies and other no-code tools to build the platforms. Agencies started at 10.000€, and other marketplace tools had far fewer features and cost around 5.000€, which made them choose Sharetribe. They needed secure payments, direct messaging between seller and buying, accounts, advanced filters, categories and video integration.

    **After launch:** With the feedback from their first users, they adjusted the available filters and categories, added more payment options and a video field and removed collaborations with ADAC and Bikefinder (checks if a bike was stolen). They also tried to acquire more stores and users. They could not add a listing view count or a feedback question about how the bike was sold when a listing gets deleted.

    **Keep or change:** They would start with Sharetribe again, as it was cheap and quick to build. They would have liked the newspaper article later, with more content on the platform. He also talked about the importance of equal commitment between the co-founders.

### 11.1.4 Aide Hulp (aidehulp14-7.com)

**Maker**: Ilse Van Dyck (2 ppl. team) | **Platform**: Softr | **Launched**: July 2021

**Type:** Marketplace | **Time to market:** 2 hours

**Description:** Aide Hulp was built in response to the floodings that happened in July 2021 in Belgium, Netherlands and Germany (Ahr Valley). It is a crisis response marketplace for the floods that happened in Belgium. People can as for and receive help from people.

**Traction:** 1,000 real-time users and 270,000 page views in the first days. 20,000 page views in the following months.

**From idea to launch:** The maker (37yo) and her friend are professionals with about years of experience in digital marketing. When the flooding happened on the 14th of July 2021, they decided to build something the next day and launched it in the afternoon. They chose the tech stack because she knew it from building a job platform for marketers in Belgium and because they did not know any other tool besides Google Sheets. It also fits their use case. It was cheap and extremely fast. They needed a solution for people to create a listing, either offering or asking for help and show these listings immediately on the platform.

    **After launch:** The demand was not anticipated. Suddenly their page was shared on radio, social media and local TV. For a short period, they acted as the local crisis relief centre as they were the only ones being so fast to act.

Soon the platform started to feel "glued together". Many people published personal data on the platform, which they solved with a big disclaimer! Since multi-language support (Dutch and French-speaking countries) was unavailable, they used a mix of both languages on the site. They also learned that people from the flooded rural regions were older and less tech-savvy than their urban non-flooded regions, which led to an oversupply of support. To keep below 10,000 Airtable rows, they sent emails asking people if the listing was still relevant to delete them manually. The need shifted from material supply to more emotional support as time passed. Hence, they later built a buddy matching system and granted direct database access to city majors for better coordination.

    **Keep or change:** Keep the tech stack and think more about the user.

## 11.1.5 Connectolingo

**[Startup and maker name is anonymised]**

**Maker**: Benjamin (9 ppl. team) | **Platform**: Bubble | **Launched**: 2018

**Type:** Web App | **Time to market:** 6 months (next to CEO full-time job)

**Description:** Connectolingo is a language-learning app for Hebrew, Arabic and English.

**Traction:** 30,000 users, 10-15% are paying.

**From idea to launch:** The maker (41yo) has an economics and accounting background and worked for a large accounting firm before founding Connectolingo 11 years ago. He started to build Connectolingo using Microsoft Silverlight and .Net. Schools and universities started to use it before Microsoft discontinued support for these platforms. They rebuilt it a second time using WordPress and Articulate Storyline (articulate.com/360) as course software. Then it took off. They could show that their language courses are a "cure for racism" in Israel – a newspaper invested in them.

Then he discovered Bubble, tried it and decided to build a new version next to the day-to-day work. They needed a mobile and web-friendly solution that integrated into their existing courseware, allowed users to pay for multiple courses and displayed the correct courses with progress to the right users. He built it in six months but would have needed only two months if he had worked full-time.

**After launch:** They changed everything since they launched Bubble, always based on user data and feedback. The landing page, the course lengths, and their free and paid model are a few examples. They had challenges in implementing different payment providers as Stripe is not available in Israel, integrating the courseware to show the correct lesson in an Iframe, and scaling the app to a high number of users. He now has the experience in Bubble to build a scalable, future-proof app that will support many more users and languages.

**Keep or change:** "I still choose Bubble because I have no regrets, maybe, you know, a product decision. But those are things that I couldn't know if I learned them without actual users, so the best thing you can do is, build an MVP. Build the product as fast as you can, and put it in front of users as fast as you can. This is the only way you can really learn what's going on, and Bubble is perfect for that. You can really worry about scaling and proper recommendations later. For starting something, it's perfect" (Connectolingo, p. 10).

### 11.1.6 Car Mingle

**[Startup and maker name is anonymised]**

**Maker**: Jan (50 ppl. team) | **Platform**: Sharetribe | **Launched**: 2019

**Type:** Marketplace | **Time to market:** 3mo. for the platform, 9mo for legal.

**Description:** Car Mingle is a peer-to-peer car-sharing platform. Anyone can provide their car to be rented by anyone. Custom hardware technology removes the need for car key handover.

**Traction:** 180,000 customers, received 4.2 million in funding and acquired a hardware startup.

**From idea to launch:** The maker has a marketing background and started Car Mingle in 2018. They asked themselves what the core functionality would be and settled on a marketplace where people could offer or rent cars, they needed a payment solution as well as a user verification mechanism and notifications. They looked into existing SaaS platforms at the time to understand which platforms come closest and explored the option of custom developing it. When they found Sharetribe, they talked with the founders, who helped them find existing Sharetribe developers that could build their first version using Sharetribe Flex (their low-code offering). They also saw other successful marketplaces using Sharetribe, which made them choose it. "If they can do it with a sauna, you can do it with a car as well". (Sharetribe, p. 2) Initially, they wanted to custom-develop a solution in parallel but stopped after two weeks as they realised it was too slow and expensive. Eventually, three Sharetribe experts developed the first version in three months for $20,000 dollar. The bottleneck for the launch was existing legislation and insurance that needed a total of 9 months until they could launch.

      **After launch:** Today, they have 180,000 customers, acquired a hardware startup that allows them to unlock cars via the air and received 4.2 million in funding. They set themselves an arbitrary number of 100,000 customers to switch to a custom solution but did not have a suitable CTO at the time. Once they found one at, 180,000 customers, thes started to rebuild it in custom-code. They hit a ceiling and wanted to improve the app's (API calls) and add specific features like location tracking, payment methods available in Singapur, social logins and more.

**Keep or change:** Start with Sharetribe again, and try to find a CTO earlier.

## 11.1.7 Paperound (paperound.com)

**Maker**: Jake Fox (4 ppl. team) | **Platform**: Sharetribe | **Launched**: January 2021
**Type:** Marketplace | **Time to market:** 1 month
**Description:** Paperound is a digital marketplace that connects small businesses with students to work on ad hoc digital projects together. Students are pre-vetted and skilled to do various kinds of tasks like content creation, research or outreach, e.g.
**Traction:** 400 taskers (students), 500+ completed projects, 50k+ in student earnings

**From idea to launch:** Jake has a BA in entrepreneurship and management and a master in economics and finance. In his first venture, he tried to build the "Strava for Tennis players" and had an app developed for him. They added too many features that no one needed. He then worked in a property startup where he had multiple marketplace ideas. "With Paperound, I had my approaches flipped. I'm only going to make as much progress as I have proof and need to build proof for myself because there's no point pouring so much time into this if I don't know if it works. So what's the quickest way to work out if it works" (Paperound, p. 1). During Christmas in 2021, he started to build Paperound, and after one week, he was done with the first version. After three weeks, he had the first taskers (students and friends of his) on board, as well as the first businesses. All on the free trial of Sharetribe.
**After launch:** It took him another two months to prove his hypothesis that businesses would return to book students. He started to raise money and got an investment of $150,000 in March. They founded the company in April 2021.
They spent quite some time changing to looks and functionality of the original Sharetribe design. They also added many features like event-based engagement (send a Whatsapp reminder if someone does not reply, e.g.). Because Sharetribe does not offer direct access to the code base on this plan, they had to build complicated automation and hacks to make it work. They used a shed and made it look like a house. This customisation was their main challenge as the tools provide a "bit of what you wanted, but it won't do everything you wanted to do" (Paperround, p. 4)
**Keep or change:** He would choose Sharetribe again to prove his MVP. Only make as much progress as you have proof. Next time he would spend more time investigating how one can interact with the platform in detail before committing to one.

## 11.1.8 Locumify

**[Startup and maker name is anonymised]**

**Maker**: Pranav (9 ppl. team) | **Platform**: Bubble | **Launched**: 2019

**Type:** Marketplace | **Time to market:** 6 months (next full-time business).

**Description:** Job platform for doctors [In one European country] to find locum work. It combines all locum job agencies [in this country] working with hospitals to offer "freelance" jobs in one platform.

**Traction:** 100,000 jobs added, 2,000+ doctors, 20 recruitment agencies that cover 90% of locum jobs [in this country], 1,1 million seed funding, profitable now.

**From idea to launch:** The maker is a trained doctor with many years of experience. When he started Locumify with his co-founder in 2016, he became the PM. They started a job platform on a WordPress website and soon custom-developed a solution with a team of engineers. He got into no-code by trying to optimise the process with Zapier, as the dev team could not ship features fast enough. They realised that their approach wasn't showing traction, and he started to experiment with a new approach using Bubble. Instead of connecting hospitals with doctors, they started to connect locum agencies that already worked with hospitals. They realised that one person had built a fully working product in less time than their four-person developer team. When the pandemic hit, they transition to a remote setup and decided to only commit to the Bubble app, which freed up an extra year of runway.

**After launch:** Once they had made the switch, they built new features. They built a compliance portal that allowed doctors to upload all their documents in one place. Although there are companies that offer this as their main product. The training navigator was a second feature - a "Glassdoor for doctors". Doctors can rate and review hospitals and see salaries. The third feature was a job recommendation engine based on 30 data points. With this feature, they pushed the boundaries of Bubble. With the Bubble team, they built an algorithm that would run through thousands of records instantaneously to suggest the right job. Each feature only took him (one person) four weeks to build. Over the years, he did struggle with building a scalable app in the beginning, like thinking in abstractions, writing clean "code, and building a functional data model. He is confident that Bubble can scale into the hundred thousand, which is enough for their product.

**Keep or change:** He would keep that a doctor can turn PM and developer.

## 11.1.9 ClimateCollabEU

**[Startup and maker name is anonymised]**

**Maker**: Tom (3 ppl. team) | **Platform**: Softr | **Launched**: Dezember 2021

**Type:** Community + Marketpl.| **Time to market:** 2 days (6 months data collection)

**Description:** European platform and community for anyone interested in climate change. Investors, employees, founders, talents and newbies.

**Traction:** 600 website signups, ~70% joined the community, 15k email subscribers, ~ 100 new/week, 500 page visits from SEO. Data that maker added: 8,000k jobs on the platform (1,000 new jobs each week), 3,000 climate startups (that maker mostly added), 4,000k investors (that maker mostly added)

**From idea to launch:** The maker (27yo) has a business background and worked in Venture capital before building ClimateCollabEU. He wanted to build his own venture and allowed himself roughly one year to make something work. Therefore he needed something where he could fail fast, make changes quickly on his own and without code. His first product was a list of climate tech startups that people could filter and see details on a separate page. He used Phantom Buster to scrape data from about 1,500 climate tech startups which took him 6 months next to his job. Building the front end for this data only took him 2 days in Softr. Anyone who signed up could add and change data. Because he saw other people who built similar communities and directories with Softr, he was confident that it was the right stack.

    **After launch:** After launch, he built an investor database based on the existing funding data from the startup database. Once he had a list, he reached out to investors to update and verify their data. He then added an employer database which essentially is a list of startups, cooperates, consultancies and VCs made for anyone looking for a job in climate tech. He also launched a job board and a community. With all these features and databases, he quickly had quite some expanses for Airtable. His number one priority was finding ways to monetise it. At the time of the interview, he experimented with a paid community and paid job posts.

    **Keep or change:** Although not a priority and the reason that allowed him to be quick, he had wished for more customisation and design flexibility and would spend more time thinking about automation and data structure.

## 11.1.10 Talent Tap

**[Startup and maker name is anonymised]**

**Maker**: Fernando (2 ppl. team) | **Platform**: Bubble | **Launched**: Jan. 2021

**Type:** Marketplace | **Time to market:** 1.5 months

**Description:** Talent Tap is a job platform where companies can poach talent based on their next desired career step. Candidates say what they are looking for in their next career step, and companies apply for the talent.

**Traction:** 11,000 talent base, 80 companies, 8 companies subscription, rest success based.

**From idea to launch:** The maker has a background in business and supply chain management and wanted to start something on his own. With his co-founder, they got accepted into an entrepreneurship program. Their CTO left the team before the program started, so they had to build a solution without writing code. They collected all the requirements and sorted them by priority using the MOSCOW framework. They loved the Lean Startup method, so they built a landing page on Webflow to test if they could get candidates for a decent use acquisition cost. He started to learn Bubble, and after 1.5 months, he learned Bubble and built the first functional version. The product needed a signup functionality to store user details like preferred job, salary etc., and recruiters needed to create campaigns. They also built a dashboard to keep track of all the KPIs. **After launch:** After launch, they grew technical debt quickly. Up until 2-4,000 users, everything worked. However, then the search function slowed down, the single-page dashboard started to load longer, and some large workflows started to break. They changed some of their backend logic and API flows and outsourced the search functionality to Agolia (an API database search tool), and he learned more about relational databases. He had the feeling that often times to "most straightforward solution was not the most scalable" in Bubble. They also completely disregarded how to match candidates automatically. They added it later and manually had to label 8,000 entries. During fundraising, they found investors, offering an existing software team. They decided to rebuild a similar solution also because providing GDPR-compliant APIs to applicant tracking software like Personio would be challenging. **Keep or change:** They would keep the development speed, quick validation of MVP with no-code and fast iteration with real users, and learn relational databases sooner.

## 11.2 No-code history timeline

**All-in-one platforms and single-function software (the 1990s)**

- 1985: Microsoft releases the first version of Excel, a tool that allows anyone to manipulate data without code (oddly, on a Mac!)
- 1993: Tim Berners-Lee invents the World Wide Web accessible to everyone, but knowledge of HTML is needed to build sites.
- 1994: Geocities releases one of the first no-code website builders, mainly hosting personal pages and fan sites.
- 1997: A trademark application is submitted for the term "cloud computing", a system that allows many no-code and low-code tools to function today.

**Extended functionality, plug-ins, and app ecosystems (the 2000s)**

- 2000: The rise of SaaS begins with Salesforce's "The End of Software" campaign.
- 2001: Mailchimp launches no-code email newsletter templates and marketing automation service.
  Enterprise low-code tools like Outsystems (2001) and Mendix (2005) launch and simplify business processes.
- 2003: Launch of WordPress, a no-code web development tool. It overtakes Geocities and soon introduces plug-ins that allow anyone to sell products online, see visitor counts and more. Other website builders like Wix, Joomla, and Squarespace create competition, but WordPress maintains its popularity.
- 2005: Salesforce App Exchange introduces the idea of plug-ins to the business world. Marc Benioff, the founder of Salesforce, famously purchased the rights for the App Store name before Apple and gifted it to Steve Jobs, his Mentor.
- 2006: Shopify (36b$ market cap in 2019) launches, a no-code e-commerce platform builder, bringing online shops to the masses.
- 2007: Launch of the iPhone and, with it, the App Store, creating the mobile app market as we know it.

**Build apps without code (the 2010s - present)**

- 2011: Zapier launches and enables non-technical people to use APIs to integrate web applications and automate workflows without writing code.
- 2012: Bubble launches its no-code web app development platform, which has become one of the most versatile solutions to date.

- 2013: Webflow launches a codeless, 100% visual way to create powerful, flexible websites and apps to remove coding as a barrier.
- 2014: Forrester coins the term "low-code" to describe development platforms that are simple, easy to use, and less dependent on coding.
- 2018 / 2020: Microsoft launches PowerApps, their in-house low-code platform, and Google acquires AppSheet, a no-code web app builder and automation platform.
-

## 11.3 No-code market overview - detail

The below selection is based on information from two no-code tool directories (*Nocodelist*, n.d.; *Nocode.Tech*, n.d.) and the Nocodeconsensus report (Bubble Group, Inc., 2020)

**Web app builder:** No-code web app platforms allow anyone to create responsive, dynamic, data-driven web apps with user logins and payments. Use cases include marketplaces, web apps, directories, websites and internal tools. Most of them offer templates that speed up development time. On the Bubble marketplace, makers can buy ready-made templates mimicking well-known platforms like Product Hunt, Facebook, Fivver, Airbnb or task managers.

- **Bubble** ([bubble.io](bubble.io)): Multi-purpose platform, higher learning curve, extremely flexible, but covers most use cases, including mobile apps.
- **Sharetribe** ([sharetribe.com](sharetribe.com)): Dedicated marketplace platform with nearly no learning curve, set up within hours, but with limited customisation abilities.
- **Softr** ([softr.io](softr.io)): The front-end for Airtable or Google Sheet data. Perfect for directories, internal tools, and simple marketplaces. Almost no learning curve and an effortless setup.
- **Webflow** ([webflow.com](webflow.com)): Design-focused front-end development platform that allows building custom websites. Paired with additional no-code backend tools like Tangram ([tangram.co](tangram.co)), this can become a flexible solution.

  **Notable mentions:** Glide ([glideapps.com](glideapps.com)), Pory ([pory.io](pory.io)), Bildr ([bildr.com](bildr.com)).

**Databases and docs:** No-code database tools provide a visual interface to manage and connect any kind of data to create one source of truth. They usually integrate well with existing no-code platforms and are used in conjunction. Recently, a new generation of docs started to integrate database features, making them more accessible than ever.

- **Airtable** ([airtable.com](airtable.com))**:** Visual database tools, a market leader with endless integrations, useful for individuals and organisations.
- **Rows** (rows.com): Next-gen spreadsheet that easily integrates many data sources.
- **Notion** ([notion.so](notion.so)): Combination of note-taking /doc and databases. Many use Notion to run websites, publish directories and build custom workspaces with superpowers.
- **Coda** ([coda.io](coda.io)): Brings words, data, and teamwork into one powerful doc. Custom buttons and formulas offer additional flexibility. It integrates well into other apps.

**Automation and API integration:** Automation platforms connect and integrate different services and platforms to make them "talk". Where previously coding was required to connect APIs, there is now a visual interface for that.

- **Zapier** ([zapier.com](zapier.com)): A market leader with 5000+ apps that can be connected.
- **Make - formerly Integromat** (make.com): Non-linear automation service.
- **IFTT** (iftt.com): Integration and automation platform focused on mobile apps and services running on Android and iOS.

**Mobile app builders:** Mobile app builders allow non-technical users to build mobile apps. They can be wrapped mobile-web apps downloadable via the browser or native apps for iOS and Android. Many app builders seem to have outdated design components, appear slow, and based on this paper's research, successful app businesses are rare.

- **Thunkable** (thunkable.com, 2015): Oldest native app builder.
- **Adalo** (adalo.com, 2018)**:** Most well-known and recommended. For native and web apps.
- **Bravo Studios** (bravostudios.com, 2019): Turns your custom Figma or AdobeXD designs into native iOS and Android apps. App builder with the biggest design flexibility.
- **Good Barber** (goodbarber.com, 2011): Design-focused native app builder with drag-and-drop components for iOS and Android.
- **Progressier** (progressier.com, 2020)**:** Turns any no-code web app into powerful progressive web apps (mobile apps) that feel like native apps.

**Enterprise low-code:** Enterprise low-code tools are tailored to large organisations and usually have integrations with enterprise SaaS like Salesforce, Oracle or SAP. Most of them were founded in the early 2000s and still solve business problems today.

Vendors include: Appian (appian.com, 1999), Outsystems (outsystems.com, 2001), Mendix (mendix.com, 2005), and Bettyblocks (bettyblocks.com, 2010).

**More categories:** For this paper, the above categories cover the primary building blocks that modern makers need to launch a software business on no-code and understand the LCNC landscape. There are more categories like website builders, eCommerce, form builders, payments, chatbots, voice assistants, marketing tools and more. (Nocodelist, n.d.; Nocode.Tech, n.d.)

## 11.4 Interview Questions Blueprint

Intro

- Say hi, mingle and ask to record the call.
- Get the agreement on video.
- The interview will be used as a case study in my bachelor thesis about no-code. (This means our conversation will be used in the thesis and also added as an appendix to the thesis.)
- We can anonymise your startup/project and name in the thesis if you want.
- We will take about one hour, maybe less, definitely not more.
- No right or wrong answers.
- Feel free to stop the interview anytime at any time.

Background and Context:

- Could you share a bit about yourself? What's your background and age?
- How did you get into no-code?
- What no-code tools do you know?
- Please describe your startup in a few sentences.
- What type of project is that? Side Hustle/Startup
- Team Size

Early Days: From idea to launch (Zero to one):

- What were your initial product requirements?
- What were your initial requirements for a no-code platform?
- Which no-code stack/platform did you use?
- Why did you choose to build and launch your product using the above no-code platform(s)?
- Did you try other solutions before? Why did you decide against them?

- Why did you decide against building a custom-coded solution?
- What challenges did you encounter while building your initial product with no-code?
- What benefits did you encounter while building your initial product with no-code?
- What was your time to market? How was that influenced by using no-code?
- What was/is your long-term product strategy?
- Rapid fire.

Later Days. From launch to scale - One to infinity):

- What challenges did or do you currently have during growth regarding the chosen no-code platform?
- What benefits do/did you encounter during startup growth regarding the chosen no-code platform?
- What changes did you perform, or features did you add after your initial product launch? What role did no-code play?
- Have you thought about re-building your product with a custom-code solution? If yes, why. If not, why?
- Did you switch or modify your tech stack after launch? If yes, why?
- What's your revenue/traction?
- Looking back, what would you change?
- Looking back, what would you keep?

Wrap up:

- Fill out the rating scale survey: Link
- Do you know anyone who built a no-code product?
- Anything you would like to add/go deeper in?

# 11.5 Detailed Figures

## 11.5.1 Figure 13: Initial requirements (detailed)

| Requirements | Softr.io | | | | Bubble.io | | | Sharetribe.io | | | Total |
|---|---|---|---|---|---|---|---|---|---|---|---|
| | Interiorly | 100Daysofn... | Aide Hulp | ClimateCollab | Connectolingo | Locumify | Talent Tap | Bike2Loop | Car Mingle | Paperound | |
| **Initial Product requirements** | | | | | | | | | | | |
| Browse Listings (Search, Cate | x | | x | x | x | x | x | x | x | x | 9 |
| Filter | x | | x | x | | x | x | x | x | x | 8 |
| Create listings | | x | x | x | | x | | x | x | | 6 |
| Login, Signup and user accounts (with info) | | | | x | x | x | x | | x | | 5 |
| Payments | x | | | | | x | | x | x | x | 5 |
| Detailed Page | | | | x | | | x | | x | | 3 |
| Mobile and Desktop friendly | | | | | x | x | | | | | 2 |
| Messaging | | | | | x | | | x | | | 2 |
| **Initial no-code platform requirements** | | | | | | | | | | | |
| Subscription /Payments | x | | | x | x | | | x | x | | 5 |
| up fast. (Easy to fugure things out with) | | | x | x | | x | | | | x | 4 |
| Cost save / cheap: | | | | | x | | | x | | x | 3 |
| freedom, no (perceived limitations) | | | | | x | x | x | | | | 3 |
| organise content | x | | x | | | | | | x | | 3 |
| and EMails) | x | x | | x | | | | | | | 3 |
| Database (Add logic) | | | x | x | x | | | x | | | 3 |
| Ability to build a fully functional marketplace add listings, and search these listings,pay for it. | | | | | | | x | x | x | | 3 |
| Decent learning curve: Detailed (Buble dev) course (from Zeroqode) | | | | | x | | x | | | | 2 |
| platform | | | | | | | | x | | x | 2 |
| ask questions to and find answers | | | | x | x | | | | | | 2 |
| User manage their own content (Login / Accounts | x | | | | x | | | | | | 2 |
| **Primary Tech Stack** | Softr + Airtable | Carrd + Mailerite (MVP) Later: | Softr, Airtable, Actito (Mail) | Softr, Airtable, Phantom Buster | Bubble, Airtable, Articulate Storyline | Bubble (App), Webflow (Website) | Bubble (App), Webflow (Website), | Sharetribe .io, Paypal, Stripe, G | Sharetribe | Sharetribe (Marketpl ace), | |
| **Product Strategy** | Yes, more | No, do small | No, timing | No, did lean | Yes, run two | Yes, use | Yes, did lean | Yes, monetise | Yes, add | No | Yes (6x) |

## 11.5.2 Figure 16: Challenges and limitations (detailed)

| Challenges and limitations | Softr.io | | | | Bubble.io | | | Sharetribe.io | | | Total |
|---|---|---|---|---|---|---|---|---|---|---|---|
| | Interiorly | 100Daysofno | Aide Hulp | ClimateCol | Connectol | Locumify | Talent Tap | Bike2Loop | Car Mingle | Paperound | |
| **Limitations & Challenges MVP** | | | | | | | | | | | |
| No limitations | x | x | x | x | | | | x | | | 5 |
| Skill of the maker (learning it/make it scalable) | | | | | x | x | x | | | | 3 |
| Customising out of the box platform | | | | | | | | x | x | x | 3 |
| Mobile responsive engine (old Bubble Engine) | | | | | x | | x | | | | 2 |
| Integrating many different services | | | | | x | | | | x | | 2 |
| UX (slow loading pages, customisation limits) | | | | | | x | | | | x | 2 |
| Lack of tutorials & ressouces (when they started) | | | | | | x | x | | | | 2 |
| Integrating payments in their country | | | | | x | | | | x | | 2 |
| **Limitations & Challenges Growth** | | | | | | | | | | | |
| Skill of the maker (Risk of building unscalable apps) | | | | | x | x | x | | | | 3 |
| Long workflows | | | | | x | | x | | | | 2 |
| Limited insights | | | | | | | | x | | x | 2 |
| Multiple payment methods | | | | | | | | x | x | | 2 |
| Design limitations / Customisation | | | | x | | | | | x | | 2 |
| Staying within Airtable limits (costs) | | | x | x | | | | | | | 2 |
| How much should they automate | | P | P | | | | | | | | 2 |
| Self user payment / subscription mgm | x | x | | | | | | | | | 2 |
| Many Customer service requests | P | | | | | | | | | | 1 |
| GDPR / Privacy (how handle user data) | | | x | | | | | | | | 1 |
| Multi language support. | | | x | | | | | | | | 1 |
| No-code solution felt "glued together" | | | x | | | | | | | | 1 |
| UX / unpolished features (blocks) | | | | x | | | | | | | 1 |
| Conditional / Sub - filters | | | | x | | | | | | | 1 |
| No advanced SEO | | | | x | | | | | | | 1 |
| Documentation (understnad what you did) | | | | | M | | | | | | 1 |
| Only maker. (risky) | | | | | M | | | | | | 1 |
| Team work on a Bubble app | | | | | M | | | | | | 1 |
| Agency set API wrong, -> slow | | | | | | | | | M | | 1 |
| Event based engagement /notifications | | | | | | | | | | x | 1 |
| **What would you change?** | | | | | | | | | | | |
| Maker activities | | | (M) | | | (M) | (M) | (M) | (M) | (M) | 6 |
| Product decisions (in hinsight) | (P) | | | (P) | P | (P) | | | | | 4 |
| Plan / know Data structure from start | | | | M | M | M | M | | | | 4 |
| Tools wise, I wouldnt change allot | x | x | | | x | x | | | | | 4 |
| No product requirements planning | | | M | | | | M | | | | 2 |
| Team (find good co-founder, CTO) | | | | | | | | M | M | | 2 |
| Strategy, Sales | P | | | | | | | | | | 1 |
| Self user payment / subscription mgm | x | | | | | | | | | | 1 |
| Deeper research in product validatio | P | | | | | | | | | | 1 |
| Build feedback mechanism into MVP | P | | | | | | | | | | 1 |
| MVP is about positioning and sales f | P | | | | | | | | | | 1 |
| Members across multiple payment platforms | | x | | | | | | | | | 1 |
| Balancing user research and just building it | | | M | | | | | | | | 1 |
| Try to find existing automations | | | | P | | | | | | | 1 |
| Monetize and test earlier | | | | P | | | | | | | 1 |
| Use a design system from start | | | | | | P | | | | | 1 |
| Write "clean code" no repeating logic | | | | | | M | | | | | 1 |
| Better research no-code platform before commiting | | | | | | | | | | M | 1 |